\documentclass[preprint,12pt]{elsarticle}

\usepackage{amssymb}
\usepackage{amsthm}

\usepackage{lineno}

\usepackage{amsmath}
\usepackage{hyperref}       
\usepackage{url}            
\usepackage{booktabs}       
\usepackage{amsfonts}       
\usepackage{nicefrac}       
\usepackage{graphicx}
\usepackage{yfonts}

\usepackage{etoolbox}
\def\*#1{\mathbf{#1}}
\def\^#1{\boldsymbol{#1}}
\def\br#1{\left[ #1 \right]}
\def\pr#1{\left( #1\right)}

\def\ang#1{\left\langle #1\right\rangle}

\def\*#1{\mathbf{#1}}
\DeclareMathOperator{\Tr}{Tr}
\newtheorem{condition}{Condition}
\newtheorem{assumption}{Assumption}

\newenvironment{assumption*}
 {\assumption}
 {\endassumption}
 
 \newenvironment{assumption**}
  {\assumption}
 {\endassumption}

\journal{Physics of Life Reviews}

\begin{document}

\begin{frontmatter}



\title{How particular is the physics of the free energy principle?}


\author[inst1]{Miguel Aguilera}

\ead{sci@maguilera.net}

\author[inst2]{Beren Millidge}
\author[inst1,inst3]{Alexander Tschantz}
\author[inst1]{Christopher L. Buckley}
\affiliation[inst1]{organization={School of Engineering and Informatics, University of Sussex},
            city={Falmer, Brighton},
            postcode={BN1 9QJ}, 
            country={United Kingdom}}

\affiliation[inst2]{organization={MRC Brain Network Dynamics Unit, University of Oxford},
            city={Oxford},
            postcode={OX1 3TH}, 
            country={United Kingdom}}
            
\affiliation[inst3]{organization={Sackler Center for Consciousness Science, University of Sussex },
            city={Falmer, Brighton},
            postcode={BN1 9QJ}, 
            country={United Kingdom}}
\begin{abstract}
The free energy principle (FEP) states that any dynamical system can be interpreted as performing Bayesian inference upon its surrounding environment.
In this work, we examine in depth the assumptions required to derive the FEP in the simplest possible set of systems -- weakly-coupled non-equilibrium linear stochastic systems. Specifically, we explore (i) how general the requirements imposed on the statistical structure of a system are and (ii) how informative the FEP is about the behaviour of such systems.
We discover that two requirements of the FEP -- the Markov blanket condition (i.e. a statistical boundary precluding direct coupling between internal and external states) and stringent restrictions on its solenoidal flows (i.e. tendencies driving a system out of equilibrium)  -- are only valid for a very narrow space of parameters. Suitable systems require an absence of perception-action asymmetries that is highly unusual for living systems interacting with an environment.
More importantly, we observe that a mathematically central step in the argument,  connecting the behaviour of a system to variational inference, relies on an implicit equivalence between the \emph{dynamics of the average} states of a system with the \emph{average of the dynamics} of those states. This equivalence does not hold in general even for linear systems, since it requires an effective decoupling from the system's history of interactions.
These observations are critical for evaluating the generality and applicability of the FEP and indicate the existence of significant problems of the theory in its current form. These issues make the FEP, as it stands, not straightforwardly applicable to the simple linear systems studied here and suggest that more development is needed before the theory could be applied to the kind of complex systems that describe living and cognitive processes.
\end{abstract}




\end{frontmatter}



\section{Introduction}


During the last decade, the `free energy principle' (FEP) has become an influential framework which aims to provide a grand theory promoting a Bayesian interpretation of living systems \cite{friston2012free, friston2013life, friston2019free}. The FEP states that any self-organizing system (i.e. any dynamical system, and therefore any living or cognitive entity) equipped with a Markov blanket -- a statistical separation between internal and external states -- can be interpreted as performing Bayesian inference upon the surrounding environment, such that its internal states come to encode probabilistic beliefs about the external environment \cite{ friston2019free, parr2020markov}. 

The core claim of the FEP is exceptionally ambitious. It implies that the dynamics of any pair of coupled systems, under specific conditions about the interaction of internal and external states, can be described as one system trying to statistically infer the states of the second system (cf an agent and its environment). This claim licenses an interpretation of the agent as performing a basic kind of Bayesian inference and encoding beliefs about the surrounding environment \cite{hohwy2016self, clark2015surfing}.
Such an equivalence could have a far-reaching influence on the study of living systems. For example, it could enable approximate calculations of the dynamics of complex systems in terms of a more tractable description of the dynamics of their sufficient statistics.
Furthermore, the FEP has been defended based on singular claims about its explanatory power, suggesting that it reveals novel insights among fundamental psychological concepts such as memory, attention, value, reinforcement, and salience \citep{friston2009free} and unifies different aspects of motor behaviour and perception, from retinal stabilization to goal-seeking \citep{friston2010action}. In addition, it has been proposed that the FEP provides a basis for integrating several general brain theories, including the Bayesian brain hypothesis, neural Darwinism, Hebbian cell assembly theory, and optimal control and decision theory \citep{friston2010free}.

The  FEP has also inspired theories such as predictive coding \cite{friston2007variational,friston2005theory,buckley2017free,millidge2021predictive} and active inference \cite{friston2015active,friston2009reinforcement,friston2017active}, which offer explanations and models of brain function and dysfunction \cite{cullen2018active} through the lens of Bayesian inference, and have become widely influential in theoretical neuroscience and beyond. For instance, predictive coding has been proposed to be a biologically plausible model of cortical function \cite{friston2005theory,buckley2017free,millidge2020predictive}, and has been applied to explain binocular rivalry \cite{hohwy2008predictive} or attention \cite{kanai2015cerebral}. Active inference, on the other hand, has seen substantial use in modelling the behaviour of human or animal subjects in various paradigms \cite{friston2015active,friston2009reinforcement}, as well as understanding rational decision-making and behavioural control \cite{tschantz2020reinforcement,da2020active,millidge2020deep,fitzgerald2015active}. Moreover, biological theories proposing that neurons \cite{parr2019neuronal}, synapses \cite{kappel2021synapse}, bacteria \cite{tschantz2020learning} or plants \cite{calvo2017predicting} could be explicitly performing Bayesian (variational) inference by minimizing free energy gradients have been proposed and justified using explicit appeals to the FEP. While such theories do not entirely depend upon the validity of the mathematical core of the FEP reviewed here, they nevertheless derive a substantial amount of their intellectual and rhetorical support from it. Thus, the fundamental validity of the mathematical framework of the FEP is of great importance to this large and rapidly increasing modelling literature.

In a vast body of work that spans over the course of about 15 years, the FEP has detailed the mathematical steps required to derive its central claims.
In the first phase, an intuitive and heuristic idea of an imperative to minimize variational free energy was developed \citep{friston2006free, friston2009free} based on the need for any recognizable `system' to maintain itself in a low entropy configuration over time against dissipative forces trying to push it towards a high entropy equilibrium state. 
Later, this heuristic argument and intuition was more formally related to concepts from stochastic thermodynamics \citep{friston2013life, friston2012free}, specifically by bringing in the Markov blanket condition and expressing the dynamics of the system in terms of a gradient descent on a variational free energy corresponding  to a generative model of the environment of the system.
Finally, the mathematical formulation and
some of its arguments have been recently refined in a new series of publications \citep{friston2019free, parr2020markov, friston2021some}, a process that is still ongoing.

\begin{figure}
\begin{center}
\includegraphics[width=9cm]{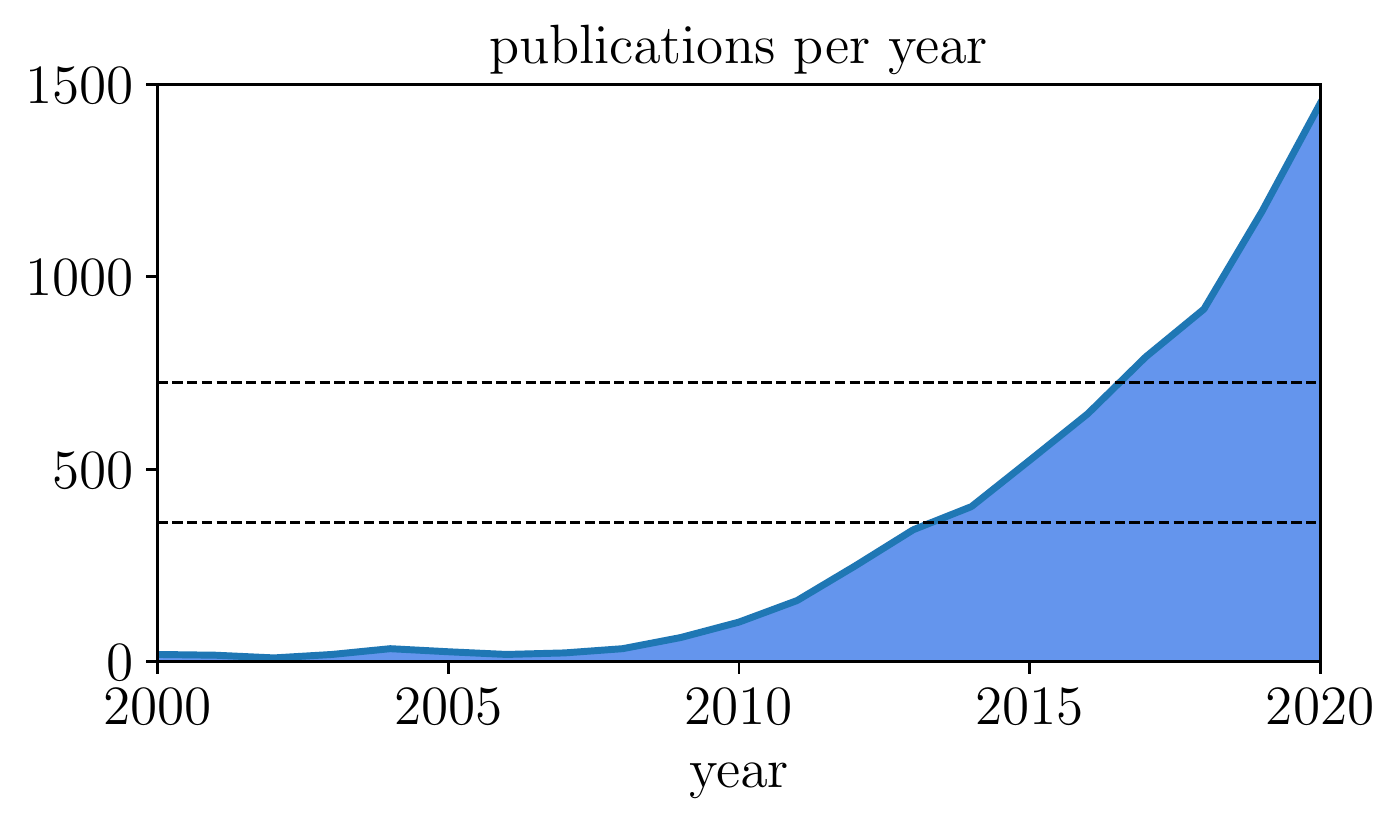}
\end{center}
\caption{\textbf{Growth in publications pertaining the free energy principle}. Publications per year searching for `free energy principle'. Source: Google Scholar. The number of papers has approximately doubled from  2013 (\href{https://scholar.google.com/scholar?hl=en&as_sdt=0\%2C5&as_ylo=2013&as_yhi=2013&q=\%22free+energy+principle\%22&btnG=}{link}) to 2017 (\href{https://scholar.google.com/scholar?hl=en&as_sdt=0\%2C5&as_ylo=2017&as_yhi=2017&q=\%22free+energy+principle\%22&btnG=}{link}) and from 2017 to 2020 (\href{https://scholar.google.com/scholar?q=\%22free+energy+principle\%22&hl=en&as_sdt=0\%2C5&as_ylo=2020&as_yhi=2020}{link}), as indicated by the dashed lines.
} 
 \label{fig:stats}
 \end{figure}

Despite the extensive literature on the FEP, there are few concrete examples that apply all the required steps to a specific, well-studied system. Similarly, it has rarely been explored whether the assumptions (which are sometimes left unstated) required for deriving the principle hold under the dynamics we expect from cognitive and living systems. 
In light of the increasing rate of publications concerning the FEP (the number of published papers on the topic doubles every few years
, Fig.~\ref{fig:stats}), we believe it is imperative to ground and test the foundations of the theory in concrete models to  assess the generality and validity of its claims.
To this end, we explore a class of systems defined by stochastic linear differential equations, under a weak-coupling assumption. 
Such systems are the simplest possible example that can display the dynamics required for the FEP, as well as capture  non-equilibrium properties of systems engaged in perception-action cycles (as we expect from living systems).
Moreover, the absence of nonlinear interactions in this class of systems allows for precise analytic calculations, which offers an interesting test-bed to examine, in detail, the connection between non-trivial dynamics and the statistical properties of coupled systems. In general, due to the special independence properties imposed by the theory, if the assumptions and steps of the FEP do not hold in such simple systems, we consider it unlikely that they hold in more complex nonlinear systems where the dynamics are expected to be more deeply intertwined. Moreover, we observe that stronger couplings result in higher-order interactions that make difficult the separation between internal and external states required by the FEP. Because of this, we expect that the introduction of non-linearities will in general have a similar effect.

\subsection*{How general is the free energy principle}

We first inspect the generality of the assumptions about the statistical structure of a system required by the principle. A crucial step to derive the FEP is to establish a relation between the average flow of change of a system that interacts with an environment and the gradient of a variational free energy of a model of this environment. 
This step relies on specific assumptions about how perception and action mediate the interaction of the internal and external states of a system. We aim to explore how general these assumptions are and whether they can be expected in the dynamical systems models of living systems.

\textbf{Perception-action interface.} The FEP partitions the states of a system into external, sensory, active and internal states. Then, the theory assumes that perception-action cycles involve causal dependencies such that internal and external states are only mutually influenced through the effect of active and sensory states \cite{friston2013life}.  We will refer to this idea as a perception-action interface (see \ref{app:definitions}).
One example of this interface is a cell membrane around a cell \cite{friston2013life}, although the formal definition of a perception-action interface does not require the interface to be a physical boundary. Thus, a perception-action interface could describe any set of variables mediating between system and environment as, for example, a combination of retinal activity and  oculomotor states mediating between neural activity in the visual cortices and the location of an object in the environment \cite{parr2020markov}.

Then, the FEP requires the system to be endowed with a particular statistical structure with two special properties: a Markov blanket (i.e. conditional independence between internal and external states) and the absence of solenoidal couplings.

\begin{figure}
\begin{center}
\includegraphics[width=13.5cm]{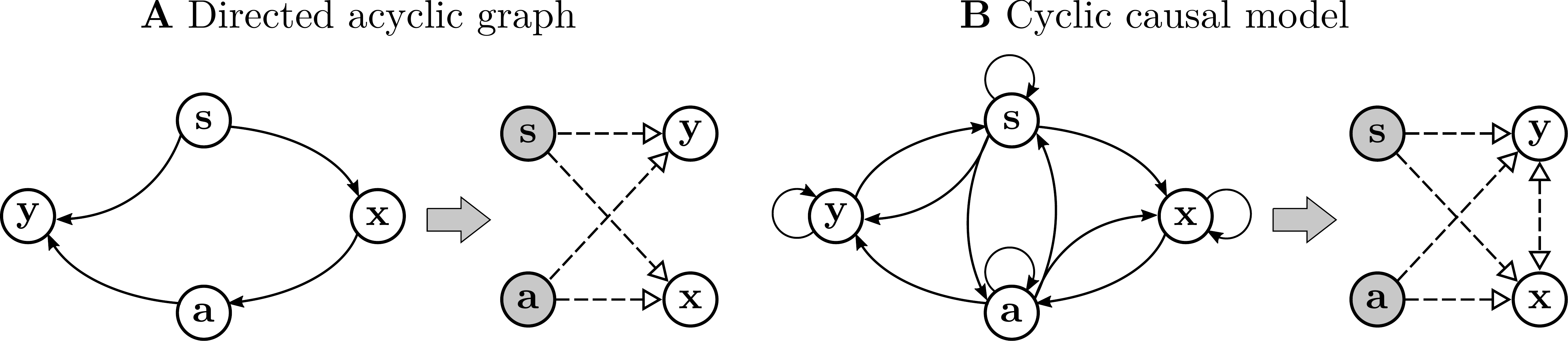}
\end{center}
\caption{\textbf{Markov blankets and causal cyclic models}. 
(\textbf{A}) An example of a Markov blanket (grey nodes $\*s,\*a$, e.g. perception and action) for a variable $\*x$ (e.g. an internal state) in a directed acyclic graph representing a Bayesian network. (\textbf{B}) An example of a cyclic causal model where the Markov blanket is not directly identifiable from the graph structure. In this case, nodes that meet the local Markov condition (grey nodes) do not guarantee the conditional independence required for a Markov blanket, as new couplings might emerge (dashed arrows between $\*x,\*y$). In each case, the graph at the right represents the equivalent correlations of the system probability distribution for fixed  $\*s,\*a$.
} 
 \label{fig:markov-blanket}
 \end{figure}

\textbf{Markov blanket.} The FEP prescribes that variables in a perception-action interface constitute a Markov blanket \citep{friston2013life, friston2021some}. A Markov blanket (see \ref{app:definitions}) is defined as a set of states (the `blanket') that separates two other sets in a statistical sense (i.e. they are conditionally independent, given the blanket). The term was initially introduced in the context of Bayesian networks or graphs \cite{pearl1988probabilistic}, and it is also known as the general Markov condition \cite{richardson1996automated}. In the FEP, Markov blankets are used for identifying a set of variables that separate the internal and external states of a system (see \cite{bruineberg2020emperor} for a detailed study on the specific use of the concept of Markov blanket in the FEP).
Here, we note that Markov blankets can be easily identified in models defined by directed acyclic  Bayesian networks (Fig. \ref{fig:markov-blanket}.A). In these systems, a sufficient condition for a Markov blanket of variable $\*x$ (e.g., an internal state) is that it contains the parent nodes of $\*x$, the children nodes of $\*x$ and the other parents of each children node (in this case, the minimum Markov blanket is composed by the grey nodes, $\*s,\*a$, in Fig. \ref{fig:markov-blanket}.A). This is defined as the local Markov condition \cite{richardson1996automated} (which implies the general Markov condition in the directed acyclic graphs of Bayesian networks).
However, in \cite{friston2013life} the FEP suggests that a Markov blanket arises naturally from the perception-action interface depicted in Fig. \ref{fig:markov-blanket}.B (although recent works restrict this to the case of an absence of particular solenoidal flows \citep{friston2021some}). However, such a cyclic structure can generate couplings that propagate beyond causal interactions. In this case, the local Markov condition does not imply a Markov blanket, as marginalization over local blanket variables generates new couplings (dashed arrows in Fig. \ref{fig:markov-blanket}.B). This important issue (identified by \citep{biehl2021technical}) contradicts the intuition that perception-action states always constitute Markov blankets, which we will explore in following sections.

\textbf{Solenoidal couplings.} The second required property is that solenoidal couplings between internal, external, sensor and motor states are absent.
The idea of solenoidal flows (see \ref{app:definitions}) arises from the separation of the flow of a dynamical system into two components. The first component is a dissipative (curl-free) flow that counters the dispersion of the density caused by random fluctuations in the system. The second component, the solenoidal flow, is defined as a conservative (divergent-free) flow capturing dissipative tendencies in the system, driving it away from equilibrium \cite[p.11]{friston2019free}.
This nonequilibrium nature is a fundamental aspect of living entities. Examples of this are asymmetric organism-environment interactions \cite{rothman1977membrane} or the oscillatory behaviour underpinning most biorhythms and neural dynamics \cite{buzsaki2006rhythms}.
The FEP assumes that  solenoidal couplings between internal and external states are absent, meaning that these flows will not penetrate through the perception-action interface.

\textbf{Average flows and the variational free energy.} 
The confluence of these properties entails an important result: due to this particular statistical structure, the average flow of a system can be  described in terms of a variational free energy gradient.
The average flow (also called the marginal flow, see \ref{app:definitions}) describes the average rate of change of the system conditioned on the blanket state. 
In turn, the variational free energy (see \ref{app:definitions}) represents an upper bound of the surprise of observed states, according to an internal model of the environment.

In the first section of this manuscript, we study how likely these conditions are for the type of dynamical interactions we expect to find in living system. By considering the simplest case of non-equilibrium stochastic dynamics, we show that Markov blankets and absence of solenoidal flows only emerge for very particular perception-action interfaces, forcing symmetries in agent-environment interactions that are not expected in living beings.

The core issue is that these three requirements -- a perception-action partition where sensor and active states mediating the internal and the external states, the existence of a Markov blanket, and decoupled solenoidal flows -- are, in principle, independent conditions. A perception-action interface does not necessarily guarantee the required conditional independence relationships and, in fact, generally does not since statistical correlations can propagate beyond this interface over time due to the intrinsic fluctuations and reentrant connections in the system. Conversely, the conditional independence relationship decreed by the Markov blanket does not imply that there is a lack of dynamical coupling between internal and external states \citep{biehl2021technical}. In practice, we discover that the kinds of systems that can fulfil both the Markov blanket condition and the block-diagonal solenoidal coupling condition are extremely specialized and generally do not possess the kind of sparsity and asymmetry of dynamical couplings that we expect from a perception-action interface.  These asymmetries are present at different levels of living systems, leading to qualitative differences between the inside and outside of a system, shaping system-environment interactions and its related flows of energy and matter \citep{rothman1977membrane, fadeel2009ins, barandiaran2009defining, ruiz2004basic}.

In other words,  although the Markov blanket assumption is typically maintained in systems with very weak couplings, we observe that there are direct interactions between external and internal states in the system's dynamics. This subtle distinction between the conditional independence relationships of the mechanisms of a system (perception-action interface) and its statistical couplings (Markov blanket) -- which is analogous to the distinction between anatomical and functional connectivity in neuroscience -- has perhaps been underappreciated in the FEP literature. This ambiguity leads to the claim that the requirements of the FEP naturally describe systems with a causal boundary between the external and internal states \cite{friston2013life}, when this is not necessarily the case.

\subsection*{How informative is the free energy principle}

Once the relation between a free energy gradient and the average flow of a system has been established,  we explore how informative this relationship is about the behaviour of an organism or the evolution of a dynamical system.

\textbf{Conditional synchronisation manifold.}
To justify the relation between a gradient of a free energy functional and the behaviour of a system, the FEP assumes the existence of a conditional synchronisation manifold (see \ref{app:definitions}). This manifold is defined as a mapping that, given a blanket state, connects the most likely internal and external states. The FEP proposes that its existence allows us to
characterise the relationship between (maximum a posteriori) internal and external states in terms of internal states `sensing' or `tracking' external states through the Markov blanket \cite{friston2019free}.

Next, the FEP links the evolution of the most likely external states with the average flow of the system, conditioned on the blanket states. 
Since the step described in the previous section connects the free energy gradient with the average flow, this assumption implies that the evolution of the most likely external states is driven by the gradient of a variational free energy. Moreover, given the conditional synchronisation manifold, the most likely internal states are also driven by this free energy gradient.
The FEP suggests this result leads to the appearance that any system with the properties described above behaves \emph{as if} internal states were performing variational inference to predict external states.

\textbf{Average flows and the rate of change of the average.} The claim that systems under the required properties behave as if performing inference relies on a venturesome assumption. It implicitly attributes the rate of change of the average (expected) state of a system-- i.e. the expected change in the internal state of an agent at a particular moment --  can be roughly described by the steady-state average flow of the system conditioned on a blanket state -- i.e. the average rate of that state during many trajectories (see \ref{app:definitions}). 
This assumption is driven by the intuition that if the average flow points in a direction that minimizes variational free energy, internal states will behave as if they are trying, on average, to minimize this free energy. As well, this can be read as an implicit assumption that these two quantities are approximately the same. However, as we will show, this intuition is incorrect since the average flow conditioned on blanket states disconnects the rate of change in the system from its previous trajectory, which is crucial to predict the system's behaviour.
In practice, this assumption implies decoupling the actions of an agent from its history of previous states. We will show that, in the class of linear system explored in this work, this results in the free energy gradients being uninformative about the behaviour of an agent or its specific trajectories.

We should note that this claim has been relaxed in more recent work \cite{friston2021some}, proposing (instead of an equivalence between rates of change of expected states and average rates) that an interpretation in terms of Bayesian inference emerges only in expectation. We will see however that this still presents important practical and conceptual problems.
To draw an analogy that portrays this issue, we could propose that the actions of a population of organisms in a particular evolutionary context maximize, on average, a fitness function -- e.g. the number of genes the population passes on to the next generation. That is, however, a largely uninformative statement for describing the behaviour of an individual organism, which depends on that organism's specific history. Moreover, the behaviour of an organism that systematically inferred what to do next in terms of (evolutionary) fitness maximization would likely be entirely different from the realized behaviour of any living organism. This is not only a point of philosophical nuance but, as we will show, translates into assumptions and important equations required for deriving the FEP.

\subsection*{Overview of the mathematical review}

In this article, we present a technical and conceptual critique of the FEP. As the theory is in continuous development, its mathematical details have been described in different forms and notations in the literature. 
For our study we try to remain as close as possible to the verbal and formal descriptions described in the most recent publications \cite{friston2019free, parr2020markov, friston2021some}. However, in at least two instances (see Assumption \ref{ass:block-diagonal-Q} and Assumptions \ref{ass:bayesian-mechanics} and \ref{ass:bayesian-mechanicsII}), we identify conflicting interpretations of the theory. In these cases, we attempt to derive our argument with as much mathematical coherence as possible while also presenting results that address the different identified possibilities. 

The rest of the article is organized as follows. First, we present a summary of the  FEP, including a list of conditions and assumptions required to derive it. Next, we survey the steps prescribed to derive the FEP for a linear stochastic system under a weak coupling assumption. We then evaluate in which cases we expect the requirements of the FEP to hold. Finally, we evaluate the implications of our study for the theory and its applicability to the type of processes that living systems are expected to manifest.

\begin{figure}
\begin{center}
\includegraphics[width=12cm]{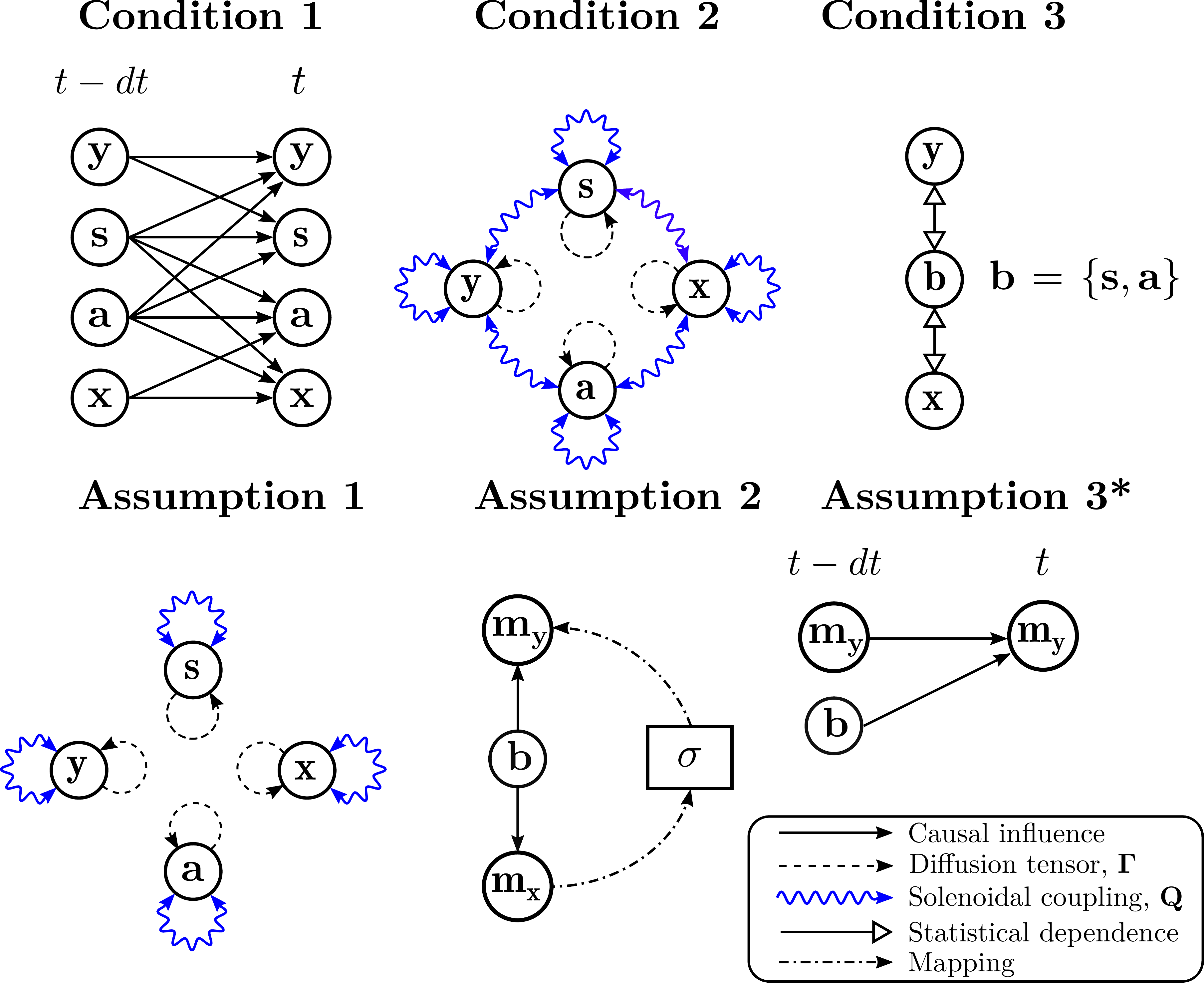}
\end{center}
\caption{\textbf{Conditions for the free energy principle}.
List of conditions and assumptions for deriving the FEP. The conditions for the FEP are: the flow of the system is constrained to define a perception-action interface (Condition \ref{cond:canonical-flow}, illustrated in the figure for an Euler integration step), the system has a global attractor described by a state-independent SDE decomposition (Condition \ref{cond:global-attractor}), and from this configuration a Markov blanket emerges (Condition \ref{cond:markov-blanket}). The assumptions necessary for deriving the FEP are: the global attractor presents an uncoupling of solenoidal flows (Assumption \ref{ass:block-diagonal-Q}), there is an invertible mapping between the most likely internal and external states (Assumption \ref{ass:sigma-mapping}) and the evolution of the most likely external dynamics is described by the marginal flow (Assumption \ref{ass:bayesian-mechanics}).
} 
 \label{fig:summary}
 \end{figure}

\section{Summary of the theory}


Here we present a succinct description of the theory, based on the most recent publications \cite{friston2019free, parr2020markov, friston2021some}, although some steps apply to previous versions as well.
Generally, the FEP assumes a random dynamical system described in terms of a Langevin stochastic differential equation:
\begin{align}
\frac{\mathrm{d}\*z_t} {\mathrm{d}t} = f(\*z_t) + \^\omega_t,
\end{align}
where $\*z_t = \{ z_{i,t}\}, i= 1 \ldots N$ is a vector, $f$ is an arbitrary but differentiable function and $\^\omega_t=\{\omega_{i,t}\}$ is a Gaussian white noise with covariance $2\^\Gamma$, which is a diagonal matrix. Throughout this article, bold symbols represent vectors and matrices.

The FEP further assumes that the system can be decomposed into external, sensory, active and internal states, $\*z=\{\*y,\*s,\*a,\*x\}$, configured as a perception-action loop reflecting an interface mediating between `autonomous' states (active and internal states $\{\*a,\*x\}$) and `non-autonomous' states (external and sensory states $\{\*y,\*s\}$). This leads to
\begin{condition}
The flow function $f$ decouples autonomous and non-autonomous states according to the following perception-action interface:
\begin{align}
f(\*z_t) =
\begin{bmatrix}
    f_y(\*y_t,\*s_t,\*a_t) \\
    f_s(\*y_t,\*s_t,\*a_t) \\
    f_a(\*s_t,\*a_t,\*x_t) \\
    f_x(\*s_t,\*a_t,\*x_t) 
\end{bmatrix}.
\end{align}
 \label{cond:canonical-flow}
\end{condition}

Under the presence of random fluctuations, some systems will converge toward a stable global attractor reflecting the steady-state dynamics of the system.
In systems out of thermodynamic equilibrium, this global attractor will describe a non-equilibrium steady-state (NESS), characterized by a continuous energy flux between the system and its environment. 
The next condition of the FEP is that this global attractor exists and can be described by a 
stochastic differential equation (SDE) decomposition \cite{kwon2011nonequilibrium,yuan2017sde} (often referred in the FEP literature as a `Helmholtz' decomposition)  describing the flow function as a linear function of the logarithmic steady state distribution of the system:
\begin{condition}
    The FEP assumes that the system will reach a non-equilibrium steady state described by the probability density function $p(\*z_t)$, which can be described using  a SDE decomposition that separates the flow into dissipative $(-\^\Gamma\nabla_{\*z} \textfrak{J}(\*z_t))$ and solenoidal $(\*Q\nabla_{\*z} \textfrak{J}(\*z_t))$ components 
    \begin{align}
        f(\*z_t) =& (\*Q - \^\Gamma)\nabla_{\*z} \textfrak{J}(\*z_t),
        \\ \textfrak{J}(\*z_t) =& -\log p(\*z_t),
    \end{align}
    where $\*Q$ is an antisymmetric matrix -- i.e. equal to its negative transpose, $\*Q=-\*Q^\intercal$. This condition requires $\^\Gamma$ and $\*Q$ to be constant matrices -- i.e. state-independent, as it is the case in linear systems\footnote{Recent works have relaxed this assumption, considering the case of  state-dependent $\^\Gamma$ and $\*Q$ matrices \citep{friston2021stochastic, parr2021memory}. However, these changes do not materially affect our critique (which uses systems where these matrices are independent of the state of the system) and we focus on the state-independent case here.}.
 \label{cond:global-attractor}
\end{condition}

The third condition is that the perception-action interface induces a Markov blanket into the NESS probability distribution. In a Markov blanket, internal states are independent when conditioned on the blanket states, composed of sensory and active states, $\*b=\{\*s,\*a\}$:
\begin{condition}
  The steady state distribution is  described in terms of a Markov blanket, where internal/external states are independent when conditioned on its blanket states.
\begin{align}
    p(\*y_t,\*x_t|\*b_t) = p(\*y_t|\*b_t) p(\^x_t|\*b_t).
\end{align}
 \label{cond:markov-blanket}
\end{condition}

It is important to note (see \cite{biehl2021technical}), that Conditions \ref{cond:canonical-flow} and Condition \ref{cond:markov-blanket} are independent and one does not entail the other. This is an important point, as some presentations of the FEP assume that a perception-action interface directly involves a causal barrier (Markov blanket) between system and environment, and this is not always the case.

\subsection{First move: Capturing Bayesian inference with an average flow}


The FEP starts by describing the average flows of the external states of a system as following a gradient minimizing a variational free energy. This connects these flows with notions from Bayesian inference.

The principle starts from a description of the `surprise' of the observed blanket states, under the steady-state distribution defined by the (random) dynamics of the system, which we denote with $\textfrak{J}(\*b)$ being the negative log-probability 
\begin{align}
	\textfrak{J}(\*b) = -\log p(\*b).
\end{align}
This implies that highly unlikely states will have a large surprise value and vice versa.

However, a system cannot access this surprise value without complete knowledge of its environment. Thus, Bayesian inference prescribes to use a lower bound of this surprise described by the variational free energy $F(\^\theta,\*b)$:
\begin{align}
	-\log p(\*b) < F(\^\theta,\*b) =& -\log p(\*b) + D_{KL}(q(\*y|\^\theta)||p(\*y|\*b))
 	\\ D_{KL}(q(\*y|\^\theta)||p(\*y|\*b)) = &  \int d\*y q(\*y|\^\theta)
 	 \log \frac{q(\*y|\^\theta)}{p(\*y|\*b)}
\end{align}
which is composed of the surprise plus a term capturing the distance from the probability of external states given the blanket $p(\*y|\*b)$ to a variational model of the environment $q(\*y|\^\theta)$ parametrized by  $\^\theta$.

This free energy constitutes a bound on the surprise, which is exact when the variational distribution $q(\*y|\^\theta)$ is equal to the reference distribution $p(\*y|\*b)$. The FEP often assumes that $q$ is a normal distribution:
\begin{align}
     q(\*y|\^\theta) =&  \mathcal{N}(\^\theta,\^\Sigma^{\theta}),
\end{align}
where $\^\theta$ are the most likely states of the system and $\^\Sigma^{\theta}$ its covariance matrix.


A simple instance of Bayesian inference can be derived from the assumption that the integral of the negative conditional log-probability (or surprise) in the Free energy equation is approximately quadratic in the region near the mode of the conditional density $\^\theta$. This is called the Laplace approximation \cite{reid2015approximate} and allows approximating 
\begin{align}
     F(\^\theta,\*b) \approx&    - \frac{n_y}{2}(1+\log(2\pi))- \frac{1}{2}\log|\^\Sigma^{\theta}| 
     \nonumber\\& + \int d\*y q(\*y|\^\theta)\pr{ \textfrak{J}(\^\theta,\*b) + \frac{1}{2} (\*y-\^\theta)^\intercal \*H_{yy}(\*y-\^\theta)}
     \nonumber\\ =&  \textfrak{J}(\^\theta,\*b)  +  \frac{1}{2} \Tr\br{\^\Sigma^{\theta}\*H_{yy}}  - \frac{n_y}{2}(1+\log(2\pi)) -\frac{1}{2}\log|\Sigma^{\theta}|.
\end{align}
where $\Tr$ is a trace operator, $\*H_{yy}=\nabla_{\*y\*y}\textfrak{J}(\^\theta,\*b) $ is the Hessian matrix respect to $\*y$ of the marginal probability distribution $p(\*y,\*b)$ at $\*y=\^\theta$.

In this scenario, finding the model $q(\*y|\^\theta)$ that minimizes the variational free energy is equivalent to approximating the distribution  $p(\*y|\*b)$. From a gradient descent perspective, if we are interested in adjusting parameters $\^\theta$, this minimization process results from following the negative gradient
\begin{align}
    \nabla_{\^\theta} F(\^\theta,\*b) =& \nabla_{\^\theta} \textfrak{J}(\^\theta,\*b).
    \label{eq:gradient_theta}
\end{align}

Under the FEP it is suggested that a system displaying a Markov blanket minimizes the free energy functional by implementing a gradient descent scheme referred to as recognition dynamics \cite{friston2009free, kim2018recognition, dacosta2021neural}. A literal interpretation of this claim involves a dynamics of the variational parameter with the form
\begin{align}
    \frac{\mathrm{d}\^\theta_t}{\mathrm{d}t} =& -\^\gamma \nabla_{\^\theta} \textfrak{J}(\^\theta_t,\*b_t),
    \label{eq:gradient-descent}
\end{align}
where $ \^\gamma$ is a matrix characterizing the rates of adjustment of the variational parameters. This type of dynamics (or a discrete counterpart) is usually proposed by active inference schemes (e.g. \cite{dacosta2021neural}). However, in the most recent articles (e.g. \cite{friston2021some}) this claim is not taken literally, and the proponents of the FEP suggest that it is only the average flows of the system (not the actual dynamics) that point in the direction of the free energy gradient.
Throughout this manuscript, we will explore both a literal interpretation of a gradient descent on the free energy and the case in which the free energy gradient is only connected with the average flows.

The FEP asserts that any system with a Markov blanket partition that reaches a non-equilibrium steady-state (NESS) can be construed as performing an elementary sort of Bayesian inference. This implies that the behaviour of a system can be described by some variable that behaves as $\^\theta_t$ in  Eq.~\ref{eq:gradient-descent}, at least on average. Specifically, the FEP proposes that the evolution of the statistics of internal states can be described in terms of the variational free energy, given a blanket $\*b_t$.

Under Condition \ref{cond:markov-blanket}, given a blanket state at time $t$ ($\*b_t$), the statistics of internal and external states can be described independently. The FEP proposes to describe change in internal and external states through variables encoding the most likely\footnote{Choosing the $\mathrm{arg\,max}$ function can imply problems in some cases (e.g. if it is non-differentiable), and in some cases the expectation has been proposed as an alternative statistic. However, in the case of Gaussian systems these two functions are equivalent, so this does not affect the conclusions in this article.} internal and external states conditioned on the blanket
\begin{align}
	\*m_x(\*b_t) =& \mathrm{arg\,max}_{\*x}  p(\^x_t|\*b_t),
	\label{eq:mode-x}
	\\ \*m_y(\*b_t) =& \mathrm{arg\,max}_{\*y}  p(\*y_t|\*b_t).
	\label{eq:mode-y}
\end{align}

Then, the average flow of the system (or marginal flow) conditioned on the blanket $\*b_t$ can be computed from the SDE decomposition in Condition~\ref{cond:global-attractor} as
\begin{align}
    \ang{ f_y(\*y_t,\*b_t)  }_{\*b_t}  =& \int d\*x_t d\*y_t  p(\*y_t,\*x_t|\*b_t) f_y(\*y_t,\*b_t) 
     \nonumber\\ =&\int d\*x_t d\*y_t  p(\*y_t,\*x_t|\*b_t) (\*Q_{yz} - \^\Gamma_{yz})\nabla_{\*z} \textfrak{J}(\*z_t)
    \nonumber\\ =&  (\*Q_{yy} - \^\Gamma_{yy}) \ang{\nabla_{\*y}  \textfrak{J}(\*z_t)  }_{\*b_t}  \nonumber\\ &+  \*Q_{yb} \ang{ \nabla_{\*b}  \textfrak{J}(\*z_t) }_{\*b_t}  +  \*Q_{yx} \ang{ \nabla_{\*x}  \textfrak{J}(\*z_t) }_{\*b_t}. 
     \label{eq:conditional-average-flow}
\end{align}



The first term in this expression can be related to the gradient of the surprise in Eq.~\ref{eq:gradient_theta} (see below).
However, the second and third terms in the Eq.~\ref{eq:conditional-average-flow} preclude a straightforward connection between the average flow of the system and the minimization of the variational free energy.
The necessary step in deriving the FEP is removing the solenoidal couplings between blocks of the system, encoded in the matrix $\*Q$, to remove these second and third terms. Thus, in order to describe the equivalence between the dynamics of a system and free energy minimization, the FEP assumes that
\begin{assumption}
Solenoidal couplings between `blocks' of states ($\*y,\*s,\*a,\*x$) are precluded when a Markov blanket emerges under sparse coupling \cite{friston2019free}.
  \begin{align}
    \*Q = 
    \begin{bmatrix}
       \*Q_{yy} 
        & 
        & 
        & 
        \\
        &\*Q_{ss} 
        & 
        & 
        \\
        & 
        &\*Q_{aa} 
        & 
        \\
        & 
        &
        &\*Q_{xx}
    \end{bmatrix}.
  \end{align}
  \label{ass:block-diagonal-Q}
\end{assumption}
This leads to 
\begin{align}
    \ang{ f_y(\*y_t,\*b_t)  }_{\*b_t} =&  (\*Q_{yy} - \^\Gamma_{yy}) \ang{\nabla_{\*y}  \textfrak{J}(\*z_t)  }_{\*b_t}
     \nonumber\\ \approx& (\*Q_{yy} - \^\Gamma_{yy}) \nabla_{\*m_y} \textfrak{J}(\*m_y(\*b_t),\*b_t).
     \label{eq:conditional-average-flow-ii}
\end{align}
Where the approximation is obtained neglecting couplings of order larger than quadratic in the average of the surprise $\textfrak{J}(\*z_t)$, as prescribed by the Laplace assumption, knowing that the flow of external states is independent of internal states $\*x$.
The  obtained expression is proportional to the gradient in Eq.~\ref{eq:gradient_theta} for $\^\theta_t = \*m_y(\*b_t)$, therefore pointing to a direction minimizing the free energy, where the factor $-(\*Q_{yy}-\^\Gamma_{yy})$ represents the rate of adjustment of variational parameters ($\^\gamma$ in Eq.~\ref{eq:gradient-descent}).

In a recent work \cite{friston2021some},  it has been proposed that in the general case $\*Q$ is only block-diagonal for autonomous  ($\{\*a,\*x\}$) and `non-autonomous' states ($\{\*y,\*s\}$), allowing  non-zero components $\*Q_{ys},\*Q_{sy},\*Q_{ax},\*Q_{xa}$. However, it is not clear (to our knowledge) how this case can be directly connected with free energy minimization, and this should be perhaps clarified by future work.
In any case, the findings in the next sections of this article bring forward similar problems considering one type of block-diagonal matrix or the other.

This step concludes the first move for deriving the FEP, by connecting the average flow of a system with the gradient minimizing a variational free energy functional. 

\subsection{Second move: Linking the average flow with the dynamics of the most likely states}

The second move for deriving the FEP involves connecting the average flow in the system with its (averaged) dynamics.
This second step starts by assuming a mapping connecting the most likely internal and external states:
\begin{assumption}
    There is a smooth and differentiable function $\sigma$ that maps between the most likely internal and external states given a blanket state,
    \begin{align}
    \*m_y(\*b_t) = \sigma(\*m_x(\*b_t) ),
    \end{align}
   and the gradient $\nabla_{\*m_x} \sigma(\*m_x)$ is invertible (i.e. $(\nabla_{\*m_x} \sigma(\*m_x))^{-1}$ exists)
   \label{ass:sigma-mapping}
\end{assumption}
A sufficient requisite for Assumption \ref{ass:sigma-mapping} is  that  the  mapping  from $\*b_t$ to $\*m_x(\*b_t)$  is  injective \cite{parr2020markov}.

Once the mapping between internal and external states is defined, the next step, as we anticipated, admits two possible interpretations. The first is an interpretation in which the dynamics of the average states of the system strictly follow a gradient descent on the free energy (in the form of Eq.~\ref{eq:gradient-descent}, e.g. \cite{dacosta2021neural}). A second interpretation relaxes this view to propose that free energy minimization only takes place on average over counterfactual trajectories (rather than directly). The distinction between the two (see Assumptions \ref{ass:bayesian-mechanics} and \ref{ass:bayesian-mechanicsII}) has been generally not discussed in detail, but it is of great importance to evaluate the claims of the FEP.

The first interpretation proposes that the dynamics of the most likely states can be described by the gradient on the free energy captured by the average flow described by Eq.~\ref{eq:conditional-average-flow-ii}, which results in:
\begin{assumption*}
   The evolution of the most likely external states is similar to their conditional marginal flow given the blanket state
   \begin{align}
       \frac{\mathrm{d}\*m_y(\*b_t)}{\mathrm{d}t} \approx & \ang{ f_y(\*y_t,\*b_t)  }_{\*b_t}  .
   \end{align}
\label{ass:bayesian-mechanics}
\end{assumption*}

The star $*$ symbol in this assumption indicates that this assumption is in general not explicitly stated, and that two competing interpretations are possible. The first interpretation, in which Assumption \ref{ass:bayesian-mechanics} holds strictly, is supported by verbal descriptions and some mathematical steps in \cite{friston2019free} and \cite{parr2020markov} (specifically the equivalents of Eq.~\ref{eq:bayesian-mechanics} and \ref{eq:derivative-mapping} below\footnote{In these works, the same symbol is used to represent $\frac{\mathrm{d}\*m_y(\*b_t)}{\mathrm{d}t} $ and $\ang{ f_y(\*y_t,\*b_t)  }_{\*b_t} $ in Eq.~\ref{eq:bayesian-mechanics} and \ref{eq:derivative-mapping} (Eq.~3.3 and 3.5 in \cite{parr2020markov}, Eq.~8.22 and 8.23 in \cite{friston2019free}), therefore making the implicit assumption that they are approximately the same quantity.
Also, it should be noted that Eq.~\ref{eq:bayesian-internal-mechanics} can only be obtained under the combination of Assumption \ref{ass:sigma-mapping} and Assumption \ref{ass:bayesian-mechanics}, as it combines properties of the mapping $\sigma$ (related to the most likely states) with the gradient of the free energy (related to the average flows).} and several verbal descriptions\footnote{Particularly sentences about variable  $\dot{\^\eta}(b)$ in these works, for which it is stated `the dynamics of the internal mode as a gradient flow on the surprisal of the external mode' (\cite{friston2019free}, p. 97) `the rate of change of the most likely internal states' (\cite{parr2020markov}), p. 7.)}).
The alternative interpretation of this assumption relaxes the equivalence between $\frac{\mathrm{d}\*m_y(\*b_t)}{\mathrm{d}t} $ and $\ang{ f_y(\*y_t,\*b_t)  }_{\*b_t} $ (see Assumption \ref{ass:bayesian-mechanicsII} below).

Taken together, these three assumptions let us derive a connection between the evolution of the most likely states and the gradient of the variational free energy:
\begin{align}
	\frac{\mathrm{d}\*m_y(\*b_t)}{\mathrm{d}t} =& (\*Q_{yy} - \^\Gamma_{yy}) \nabla_{\*m_y} \textfrak{J}(\*m_y(\*b_t),\*b_t).
	\label{eq:bayesian-mechanics}
\end{align}
This has an important implication, as it allows us to derive that $\*m_y(\*b_t)$ behaves as the variational inference parameter $\^\theta$ in Eq.~\ref{eq:gradient_theta}. Therefore, the dynamics of a system can be described as if performing variational inference.

This is possible because the conditional synchronisation manifold (Assumption \ref{ass:sigma-mapping}) allows deriving a mapping between the evolution of the most likely states through the chain rule:
\begin{align}
    \*m_y(\*b_t) = \sigma(\*m_x(\*b_t) ) \Rightarrow \frac{\mathrm{d}\*m_y(\*b_t)}{\mathrm{d}t} = \nabla_{\*m_x} \sigma \frac{\mathrm{d}\*m_x(\*b_t)}{\mathrm{d}t} .
    \label{eq:derivative-mapping}
\end{align}

Finally, under Assumption \ref{ass:bayesian-mechanics}, the dynamics of the most likely internal states can be described as minimizing the variational free energy about external states
\begin{align}
	\frac{\mathrm{d}\*m_x(\*b_t)}{\mathrm{d}t}  =& (\nabla_{\*m_x}\sigma) ^{-1}  \frac{\mathrm{d}\*m_y(\*b_t)}{\mathrm{d}t}
	\nonumber\\ \approx& (\nabla_{\*m_x}\sigma) ^{-1} (\*Q-\^\Gamma) (\nabla_{\*m_x}\sigma) ^{-1} \nabla_{\*m_x}  \textfrak{J}(\sigma(\*m_x(\*b_t)),\*b_t) ,
     \label{eq:bayesian-internal-mechanics}
\end{align}
which corresponds to following the gradient descent on the variational free energy described by Eq.~\ref{eq:gradient-descent}, now rewritten in terms of the most likely internal states.

Here, we shall note that, in general, Assumption \ref{ass:bayesian-mechanics} does not hold for most dynamical systems with stochastic fluctuations, as it equates the rate of change of an average with the average of the rate of change. 
The proponents of the FEP in recent works offer instead a more relaxed interpretation of the gradient descent on the variational free energy. This interpretation proposes that the FEP applies just to the marginal flows, and thus a system behaves `as if' performing Bayesian inference just on average. For example in \cite{friston2021some} the authors propose that `the  interpretation  in  terms  of  Bayesian inference emerges only in expectation –- or on average [\dots] The classical example here is the averaging of multiple responses to sensory perturbations, when characterizing  evoked  responses  in  internal  states'. This results in the substitution of Assumption \ref{ass:bayesian-mechanics} by
\setcounter{assumption}{2}
\begin{assumption**}
If the conditional average flows follow the direction of a descending gradient of a variational free energy, the behaviour of the states of the system can be interpreted `as if' they were, on average, performing a gradient descent or minimizing a free energy functional.
\label{ass:bayesian-mechanicsII}
\end{assumption**}

In general, it is not easy to distinguish in the FEP literature when Assumption \ref{ass:bayesian-mechanics} or Assumption \ref{ass:bayesian-mechanicsII} is considered, as verbal descriptions sometimes refer to dynamics of most likely states and average flows indistinctly. 
Note also that some important steps like the chain rule in Eq.~\ref{eq:derivative-mapping} can only be derived for the interpretation promoted by Assumption~\ref{ass:bayesian-mechanics}.
Despite the interpretation, under these assumptions the proponents of the FEP conclude that the  dynamics of the most likely internal and external states can be described as following a negative free energy gradient, which is equivalent to stating that they evolve \emph{as if} performing a Bayesian inference.
Moreover, the FEP proposes that this can be extended to the dynamics of actions $\*a$ \cite{parr2020markov}, deriving the principle of active inference \cite{friston2007variational, buckley2017free}.

The two moves described here make important assumptions about the underlying dynamical systems used to derive the theory. The first move assumes a very specific statistical structure in which a gradient of the free energy functional is directly connected to average flows in the systems, without justifying to what extent it can be expected from the classes of systems capturing properties from biological systems. 
The second move makes further assumptions to justify that average flows (i.e. the average of the rate of change) in the system are informative about its behaviour and dynamics (i.e. the rate of change of the average), supporting an interpretation that described the behaviour of a system \emph{as if} performing Bayesian inference. In the next sections, we will see that the steps for deriving this interpretation are not justified for the non-equilibrium linear systems studied in this paper.

In the rest of the document, we will take Conditions \ref{cond:canonical-flow} and \ref{cond:global-attractor} for granted and explore the generality of the other conditions and assumptions in linear stochastic systems with weak couplings.
We will see that Assumption \ref{ass:sigma-mapping} hold under special conditions and can be expected for a given class of systems. In contrast, we show that Condition \ref{cond:markov-blanket} and Assumption \ref{ass:block-diagonal-Q} only hold for very specific sensorimotor loops and Assumption \ref{ass:bayesian-mechanics} and \ref{ass:bayesian-mechanicsII} do not hold in general, threatening the viability of the FEP and its applicability to most living systems.

\section{Mathematical review of the FEP under linear stochastic dynamics}

In order to explore the assumptions enumerated above in a class of tractable non-equilibrium dynamical systems, we restrict our analysis to the class of systems captured by a linear Langevin dynamics (which can be seen as an approximation of an Ornstein-Uhlenbeck process \citep{vatiwutipong2019alternative}) defined by
\begin{equation}
    \frac{\mathrm{d}\*z_t}{\mathrm{d}t} = \*J ( \*z_t -\^\rho)+  \^\omega_t,
\end{equation}
where $\*J$ is an $n\times n$ invertible real matrix, $\^\rho$ is an $n$ dimensional real vector, $\^\omega_t$ is a standard n-dimensional Gaussian white noise with a diagonal covariance matrix $2\^\Gamma $. The linearity of the process guarantees that the model will eventually result in a Gaussian distribution.

The solution of this system (see \ref{app:linear-langeving-dynamics}) in the non-equilibrium steady state (NESS) takes the form of a multivariate Gaussian distribution $\mathcal{N}(\^\rho,\^{\Sigma}^* )$ \cite{vatiwutipong2019alternative, godreche2018characterising} with  statistical moments:
 \begin{align}
     \lim_{t\to\infty} \*m_{t} =& \^\rho,
     \\ \lim_{t\to\infty} \^\Sigma_{t} =& \^{\Sigma}^*, \qquad  \*J \^{\Sigma}^* +\^{\Sigma}^*  \*J^\intercal + 2\^\Gamma = \*0,
     \label{eq:NESS-solution}
 \end{align}
 where $\^{\Sigma}^*$ can be found numerically by solving the above \emph{continuous Lyapunov} equation.
If $\*J$ is symmetric, the steady state of the system is a state of equilibrium with $\^{\Sigma}^* = - \*J^{-1}\^\Gamma$. However, the FEP focuses instead on NESS, which are more appropriate for describing living systems.
 
For studying the NESS of the system, in this article we explore the case in which non-diagonal couplings in $\*J$ are small, although we explore the effect of considering higher orders of this approximation. Thus, we define  the coupling matrix as, 
 $\*J =  - \*I + \*C$, where $\*C^2$ are assumed to be small.
 
This leads to the following power series expansion,
 \begin{align}
     \^{\Sigma}^*  =  \^\Gamma + \frac{1}{2}(\*C\^\Gamma+\^\Gamma\*C^\intercal) + \frac{1}{4}\pr{\*C^2\^\Gamma + 2\*C\^\Gamma \*C^\intercal + \^\Gamma(\*C^\intercal)^2 } + \mathcal{O}(\*C^3).
     \label{eq:covariance-expansion}
\end{align}
 To further simplify things, we also define a homogeneously distributed noise $\^\Gamma =\varsigma^2 \*I$, being $\varsigma$ a scalar constant.
The details of the derivation of the equations above are described in \ref{app:linear-langeving-dynamics}.

\begin{figure}
\begin{center}
\includegraphics[width=13cm]{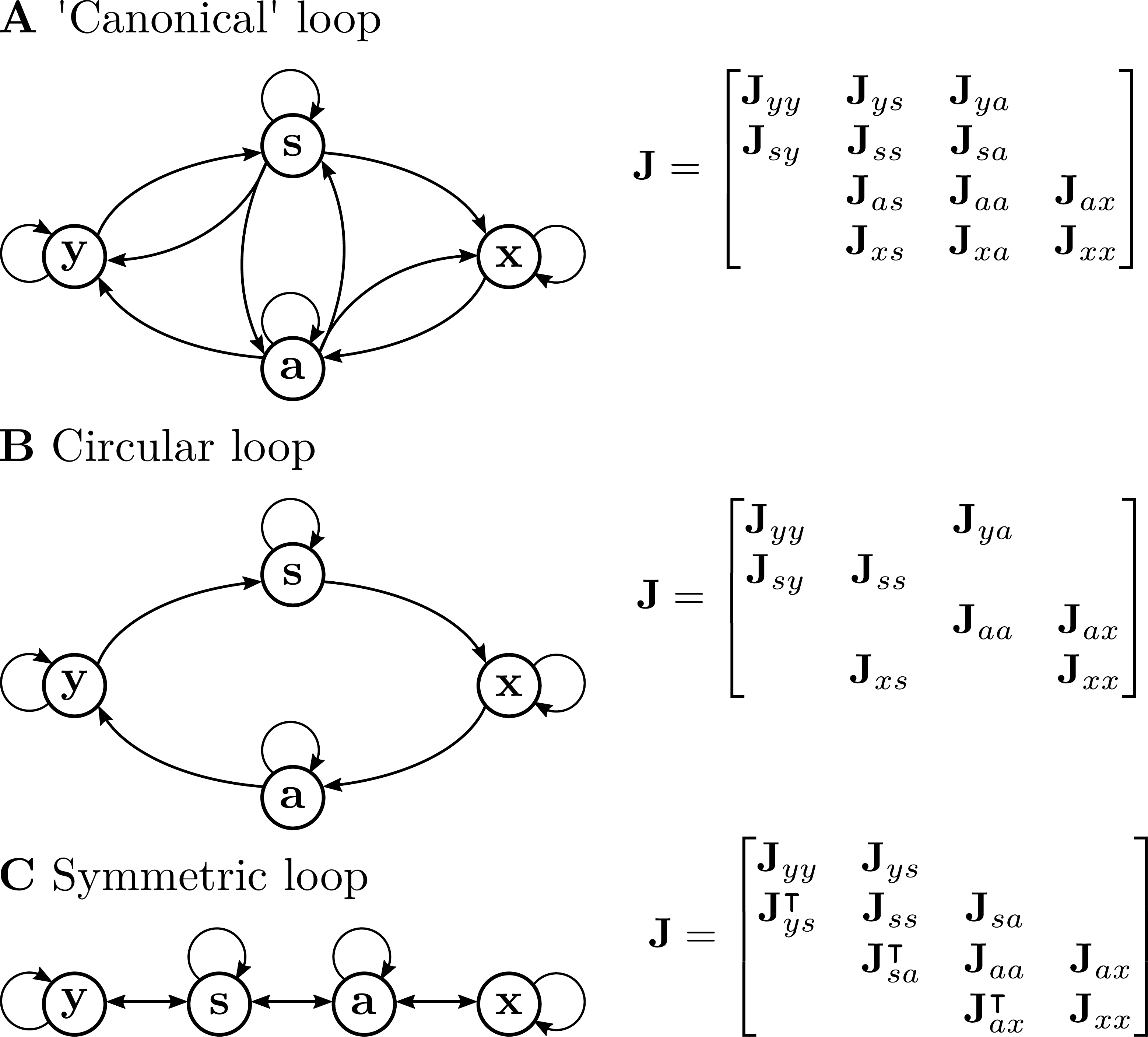}
\end{center}
\caption{\textbf{Sensorimotor loop structures}. 
The figure shows the `canonical' or general sensorimotor structure proposed by the FEP \cite{friston2021some} (\textbf{A}),  a circular loop where all connections between elements are asymmetric and unidirectional (\textbf{B})  and a restricted sensorimotor loop where all interactions between blocks of the system are symmetric, only allowing asymmetric connections within blocks (\textbf{B}).
} 
 \label{fig:sm-loops}
 \end{figure}
 
\subsection{Can we expect the requirements for deriving the FEP in living systems?}
\label{sec:requirements}

To explore the generality of the conditions required to connect the average flow of a system to the gradient of the free energy, here we explore Conditions \ref{cond:markov-blanket} and Assumption \ref{ass:block-diagonal-Q} required to derive the first move of the FEP. 

\subsubsection*{Condition \ref{cond:markov-blanket}: Markov blanket}

First, the FEP requires the existence of a Markov blanket imposing a conditional independence between internal and external states given a blanket state (Condition \ref{cond:markov-blanket}, \cite{friston2019free, parr2020markov}). We will see that not all linear systems meet this condition, although it can be considered as an approximation in the case of very weak-couplings.

For systems represented by Gaussian distributions, the Markov blanket condition (Condition \ref{cond:markov-blanket}) is  only met when the inverse of the covariance, the Hessian matrix $\*H = \pr{\^{\Sigma}^*}^{-1}$, satisfies,
\begin{align}
    \*H_{yx} = \*H_{xy} = \*0.
\end{align}
Thus this begs the questions: How common are Markov blankets? And when can we expect to find them in living systems? The FEP generally proposes that the theory holds when the Markov blanket condition is satisfied under a `canonical flow constraint' \cite{friston2021some} that is defined in our linear system by $\nabla_{\*z} f(\*z) =  \*J$ structured as in Fig.~\ref{fig:sm-loops}.A.
This means that, mechanistically, external and sensory states do not depend on internal states, and that action and internal states do not depend on external states. This is representative of the kind of asymmetries we expect from a sensorimotor loop in living systems. However, even in linear systems, it is not easy to know  under what conditions a system can satisfy both the canonical flow constraint and possess a Markov blanket. As \cite{biehl2021technical} points out, neither one of these conditions is sufficient to guarantee the other.

In the case of weak couplings where $\*J=-\*I+\*C$, and homogenous noise $\^\Gamma = \varsigma^2 \*I$, the covariance of the system can be expanded as Eq.~\ref{eq:covariance-expansion},
and the inverse covariance (Hessian) can be computed as a Neumann series (\ref{app:linear-langeving-dynamics}),
\begin{align}
    \*H =& \pr{ \^{\Sigma}^*}^{-1} =
    \varsigma^{-2} \*I -\frac{\varsigma^{-2}}{2}(\*C+\*C^\intercal)  + \frac{\varsigma^{-2}}{4}\pr{ \*C^\intercal \*C - \*C \*C^\intercal }  + \mathcal{O}(\*C^3).
     \label{eq:hessian-expansion}
\end{align}

Under the couplings determined by Condition \ref{cond:canonical-flow} we can see that for a first order approximation $	\*H_{yx} =\*0  + \mathcal{O}(\*C^2)$, satisfying the Markov blanket condition. 

For a second order weak coupling approximation we have,
\begin{align}
	\*H_{yx} = - \frac{\varsigma^{-2}}{4}\pr{ \*C_{ys}  \*C_{xs}^\intercal +  \*C_{ya}  \*C_{xa}^\intercal }   + \mathcal{O}(\*C^3).
\end{align}
Under the canonical flow constraints, relatively few systems will display an exact Markov blanket, except for combinations of parameters that happen to cancel the terms in the equation above.
One exception is systems with weak couplings under circular loops (Fig.~\ref{fig:sm-loops}.B) or systems with two layers of blanket states (e.g. the system in Fig.~\ref{fig:sm-loops}.C). Note that this is because cycles generating conditional couplings between $\*x,\*y$ are of order higher than 2. In general, these cases will not display a Markov blanket for stronger couplings (see Eq.~\ref{eq-app:hessian-expansion},\ref{eq-app:hessian_yx-expansion}, except for perfectly symmetric couplings). Thus, we can conclude that Markov blankets will emerge only for particular combinations of parameters, as cycles in the system will in general introduce couplings preventing their existence.

\subsubsection*{Assumption \ref{ass:block-diagonal-Q}: solenoidal coupling}
\label{sec:solenoidal-coupling}
 
The FEP requires that the averaged marginal flow of external states $\*y$, given a blanket state $\*b$, depends only on the gradients of its marginal density. For this, it is a requirement that there is no solenoidal coupling between external and other states (Assumption \ref{ass:block-diagonal-Q}).
We will see in this section that most linear systems will not meet this condition.

We can rewrite Eq.~\ref{eq:NESS-solution} to express the matrix of solenoidal couplings $\*Q$ as (see \ref{app:solenoidal-flow} or \cite{biehl2021technical})
\begin{align}
    \*J \*Q  + \*Q \*J^\intercal = \*J \^\Gamma  - \^\Gamma \*J^\intercal.
\end{align}

Again, the values of $\*Q$ can be obtained by solving the corresponding continuous Lyapunov equation.
Assuming  $\*J = \*C - \*I$, this matrix can be expressed as
the power series
\begin{align}
      \*Q  =  \frac{1}{2}(\*C \^\Gamma  - \^\Gamma \*C^\intercal) + \frac{1}{4}\pr{\*C^2 \^\Gamma - \^\Gamma (\*C^2)^\intercal } + \mathcal{O}(\*C^3).
\end{align}

The first order approximation under weak coupling $\*J=-\*I+\*C$ and $\^\Gamma = \varsigma^{2} \*I$ and  the canonical flow constraints (Condition \ref{cond:canonical-flow}, Fig.~\ref{fig:sm-loops}.A) results in the solenoidal coupling matrix
\begin{align}
\*Q = \frac{\varsigma^{2}}{2}
\begin{bmatrix}
   \*C_{yy} - \*C_{yy}^\intercal 
    & \*C_{ys} -  \*C_{sy}^\intercal 
    & \*C_{ya} 
    & 
    \\\*C_{sy}  - \*C_{ys}^\intercal 
    &\*C_{ss} - \*C_{ss}^\intercal
    & \*C_{sa} -  \*C_{as}^\intercal
    & -\*C_{xs}^\intercal
    \\-\*C_{ya}^\intercal
    & \*C_{as} - \*C_{sa}^\intercal 
    &\*C_{aa} - \*C_{aa}^\intercal
    & \*C_{ax} -  \*C_{xa}^\intercal
    \\
    & \*C_{xs} 
    & \*C_{xa} - \*C_{ax}^\intercal 
    &\*C_{xx} - \*C_{xx}^\intercal
\end{bmatrix} + \mathcal{O}(\*C^2).
\end{align}
Where Assumption \ref{ass:block-diagonal-Q} is not met for most parameter combinations.

In the best-case scenario, we can make many terms in the matrix above disappear by making blocks symmetric when possible. This however does not completely remove non-diagonal blocks, and leaves
\begin{align}
\*Q = \frac{\varsigma^{2}}{2}
\begin{bmatrix}
   \*C_{yy} - \*C_{yy}^\intercal  & 
    & \*C_{ya} 
    & 
    \\  
    & \*C_{ss} - \*C_{ss}^\intercal 
    & 
    &-\*C_{xs}^\intercal
    \\-\*C_{ya}^\intercal
    & 
    & \*C_{aa} - \*C_{aa}^\intercal 
    & 
    \\
    & \*C_{xs} 
    &  
    & \*C_{xx} - \*C_{xx}^\intercal 
\end{bmatrix} + \mathcal{O}(\*C^2).
\end{align}
Thus we observe that, even in a very weakly coupled system, the only way of setting $\*Q_{ya} \*Q_{sx}$ and their antisymmetric counterparts to zero is to effectively decouple some parts of the sensorimotor loop completely that results in a `symmetric' interaction loop with the form displayed in Fig.~\ref{fig:sm-loops}.C).
In this type of system, detailed balance is only broken by interactions inside blocks $\*y,\*s,\*a,\*x$, as all couplings between blocks are symmetric. Therefore, the system is driven out-of-equilibrium only by internal tendencies of these blocks, not by their interactions between them. This precludes for example the existence of asymmetric agent-environment interactions, which may be crucial for many living processes as a mechanism for generating qualitative differences between the inside and the outside of a system, as well as for regulating exchanges of matter and energy with the environment  \citep{rothman1977membrane, fadeel2009ins, barandiaran2009defining, ruiz2004basic}.




For a second order approximation, we find the solenoidal coupling terms between external and autonomous states:
\begin{align}
    \*Q_{ya} =& \frac{\varsigma^{2}}{4}\pr{2\*C_{ya} + \*C_{yy}\*C_{ya} + \*C_{ys}\*C_{sa}  + \*C_{ya}\*C_{aa} - \*C_{sy}^\intercal\*C_{as}^\intercal} + \mathcal{O}(\*C^3)
    \label{eq:Q_ya},
    \\ \*Q_{yx}  =& \frac{\varsigma^{2}}{4}\pr{ \*C_{ya}\*C_{ax} -   \*C_{sy}^\intercal\*C_{xs}^\intercal} + \mathcal{O}(\*C^3),
     \\ \*Q_{sa}=& \frac{\varsigma^{2}}{4}\Big(2(\*C_{sa} - \*C_{sa}^\intercal) + \*C_{sy}\*C_{ya} + \*C_{ss}\*C_{sa}  + \*C_{sa}\*C_{aa}
     \nonumber\\ & -   \*C_{ss}^\intercal\*C_{as}^\intercal -  \*C_{as}^\intercal\*C_{aa}^\intercal -  \*C_{xs}^\intercal\*C_{ax}^\intercal\Big) + \mathcal{O}(\*C^3),
    \\ \*Q_{sx}  =& \frac{\varsigma^{2}}{4}\pr{-2\*C_{xs}^\intercal + \*C_{sa}\*C_{ax} -   \*C_{ss}^\intercal\*C_{xs}^\intercal -  \*C_{as}^\intercal\*C_{xa}^\intercal -  \*C_{xs}^\intercal\*C_{xx}^\intercal} + \mathcal{O}(\*C^3).
     \label{eq:Q_sx}
\end{align}
We can see that this matrix will in general only be block diagonal in very specific cases, exemplifying how the presence of higher-order terms complicates the uncoupling of solenoidal terms. 

These results show how, in the case of linear, weakly-coupled systems, the assumptions about the statistical structure of a system required by the FEP is restricted to a very narrow space of parameters (i.e. values of $\*C$). In particular, removing solenoidal couplings between blocks of the system requires a highly symmetric coupling structure (Fig.~\ref{fig:sm-loops}.C). This prevents the application of the theory to many common structures found in biological systems.

\subsection{Can the FEP explain and describe the behaviour of living systems?}

Above, we have shown that the conditions for connecting the conditional average flow with the gradient of the free energy functional only hold for a very narrow class of systems. Now we explore, in the cases where the previous requirements are met, the connection between the conditional average flow and the behaviour of the most likely states of the system. That is, are the results of the FEP able to describe, explain or predict the dynamics of the systems that conform to its assumptions? Here we will see that, while Assumption \ref{ass:sigma-mapping} follows from a broad class of linear systems, one of the major findings of our study is that Assumptions \ref{ass:bayesian-mechanics} and \ref{ass:bayesian-mechanicsII} present important issues that prevent the results of the FEP to be good descriptions of the behaviour of stochastic dynamical systems

\subsubsection*{Assumption \ref{ass:sigma-mapping}:  Mapping between internal and external statistics}
 
A further requirement of  the  FEP is that a mapping $\sigma$ exists between internal and external states (Assumption \ref{ass:sigma-mapping}). We will see that in linear systems, at least, there are many systems that meet this condition.

Given a NESS well characterized by Gaussian distributions  over states $p(\*z_t) = \mathcal{N}(\^\rho,\^{\Sigma^*})$, the conditional distributions given a particular blanket state $\*b=\{\*s,\*a\}$ results in the most likely states (Eq. \ref{eq:mode-x} and \ref{eq:mode-y}) being described as
 \begin{align}
     \*m_{x}(\*b_t) =&   \^\rho_x + \^{\Sigma}^*_{xb} \pr{\^{\Sigma}^*_{bb}}^{g} ( \*b_t - \^\rho_b) ,
     \\ \*m_{y}(\*b_t) =&    \^\rho_y + \^{\Sigma}^*_{yb} \pr{\^{\Sigma}^*_{bb}}^{g} ( \*b_t - \^\rho_b),
     \label{eq:conditional-external-dynamics}
 \end{align}
 where $\pr{\^{\Sigma}^*_{bb}}^{g}$ is a generalized inverse matrix (which is equivalent to the inverse for nonsingular matrices).
 If $\^{\Sigma}^*_{xb}$ and $\^{\Sigma}^*_{bb}$ are nonsingular (this implies that $n_x\geq n_b$), we can derive the linear mapping,
  \begin{align}
     \*m_{y}(\*b_t) 
     =&  \^\rho_y + \^{\Sigma}^*_{yb} \pr{\^{\Sigma}^*_{xb}}^{-1} \pr{\*m_{x}(\*b_t) -\^\rho_x} \equiv \sigma(\*m_x(\*b_t)),
 \end{align}
 which is invertible if $\^{\Sigma}^*_{yb}$ is non-singular (for this $n_y\geq n_b$ is required)
\begin{align}
     \*m_{x}(\*b_t) =&  \*m_x + \^{\Sigma}^*_{xb} \pr{\^{\Sigma}^*_{yb}}^{-1} \pr{\*m_{y}(\*b_t) -\^\rho_y} \equiv \sigma^{-1}(\*m_y(\*b_t)).
\end{align}
 
Yielding the evolution of the most likely states as,
 \begin{align}
     \frac{\mathrm{d}\*m_{x}(\*b_t)}{\mathrm{d}t} =&  \nabla_{\*m_y} \sigma^{-1}(\*m_{y}(\*b_t)) \frac{\mathrm{d}\*m_{y}(\*b_t)}{\mathrm{d}t} =  \^{\Sigma}^*_{xb} \pr{\^{\Sigma}^*_{yb}}^{-1}  \frac{\mathrm{d}\*m_{y}(\*b_t)}{\mathrm{d}t} .
     \label{eq:sigma_gradient}
 \end{align}
 
This shows that, in general, a mapping between the most likely internal and external states  exists if the corresponding covariance submatrices are invertible.

The FEP often states that the existence of this mapping is a consequence of a Markov blanket \cite{parr2020markov}, but we note that the existence of such a mapping is independent of Conditions \ref{cond:canonical-flow} and \ref{cond:markov-blanket}.
The existence of a Markov blanket implies a conditional covariance $\^\Sigma_{yx}(\*b_t)=\*0$, but this does not affect the relation between internal and external states. Thus, in linear systems any variable can potentially mediate an invertible mapping, independently of a Markov blanket, if the corresponding covariance submatrices are nonsingular\footnote{At the moment of writing this manuscript, we have found that a similar result has been found independently in unpublished work \cite{dacosta2021bayesian}.}.

\subsubsection*{Assumption \ref{ass:bayesian-mechanics}: The  dynamics and Bayesian inference}
\label{sec:ass-3}
The final assumption for deriving the FEP is that the evolution of the most likely external states is linked with the evolution of the marginal flow conditioned on blanket states $\*b_t$.
In this section, we see that this assumption does not hold in general.

The time derivative of $\*m_y(\*b_t)$ yields,
\begin{align}
     \frac{\mathrm{d}\*m_y(\*b_t)}{\mathrm{d}t} =& \^{\Sigma}^*_{yb} \pr{\^{\Sigma}^*_{bb}}^{g}  \frac{\mathrm{d}\*b_t}{\mathrm{d}t}
	 = \^{\Sigma}^*_{yb} \pr{\^{\Sigma}^*_{bb}}^{g}\pr{ \*J_{bz}(\*z_t-\^\rho) +  \^\omega_{b,t}}.
	\label{eq:most-likely-state-evolution}
\end{align}
 
Conversely, the equation proposed by the FEP uses the marginal flow (Assumption \ref{ass:bayesian-mechanics}, see \cite{parr2020markov} or Appendix B in \cite{friston2019free}). We represent the dynamics of a variable driven by this marginal flow as
\begin{align}
     \frac{\mathrm{d}\*{\widetilde m}_y(\*b_t)}{\mathrm{d}t} \equiv &  \ang{f_y(\*y_t,\*b_t)  }_{\*b_t}  =
      (\*Q_{yz} -\^\Gamma_{yz}) \ang{\nabla_{\*z}
      \textfrak{J}(\*z_t)}_{\*b_t}
      \label{eq:marginal-flow-approximation}
      \nonumber\\ =& \*J_{yy} (\*m_y(\*b_t)-\^\rho_y) + \*J_{yb} (\*b_t-\^\rho_b),
\end{align}
where $\*{\widetilde m}_y(\*b_t)$ captures the evolution of the most likely dynamics in a system behaving in a way that strictly minimizes the free energy.
The difference between the behaviour of $\*m_y(\*b_t)$ and $\*{\widetilde m}_y(\*b_t)$ allows us to evaluate how informative is the FEP about the behaviour of a system.
If a system approximately follows a free energy gradient, the behaviour of these two variables will present some similarities in their evolution.
This approximation is crucial for the FEP as it is this marginal flow that is connected with the gradient of the free energy (therefore equating dynamics and inference, see Assumption \ref{ass:bayesian-mechanics}).
By using Eq.~\ref{eq:conditional-external-dynamics} this expression can be rewritten simply as,
\begin{align}
     \frac{\mathrm{d}\*{\widetilde m}_y(\*b_t)}{\mathrm{d}t} =& \pr{\*J_{yy} \^{\Sigma}^*_{yb} \pr{\^{\Sigma}^*_{bb}}^{g} + \*J_{yb} }( \*b_t - \^\rho_b).
\end{align}
We can thus see that these two equations represent quite different quantities and that, in general, Assumption \ref{ass:bayesian-mechanics} will not hold for linear systems.


\begin{figure}
\begin{center}
\begin{tabular}{cc}
    \multicolumn{1}{l}{\textbf{A}}  & \multicolumn{1}{l}{\textbf{B}}  \\
     \includegraphics[width=6.4cm]{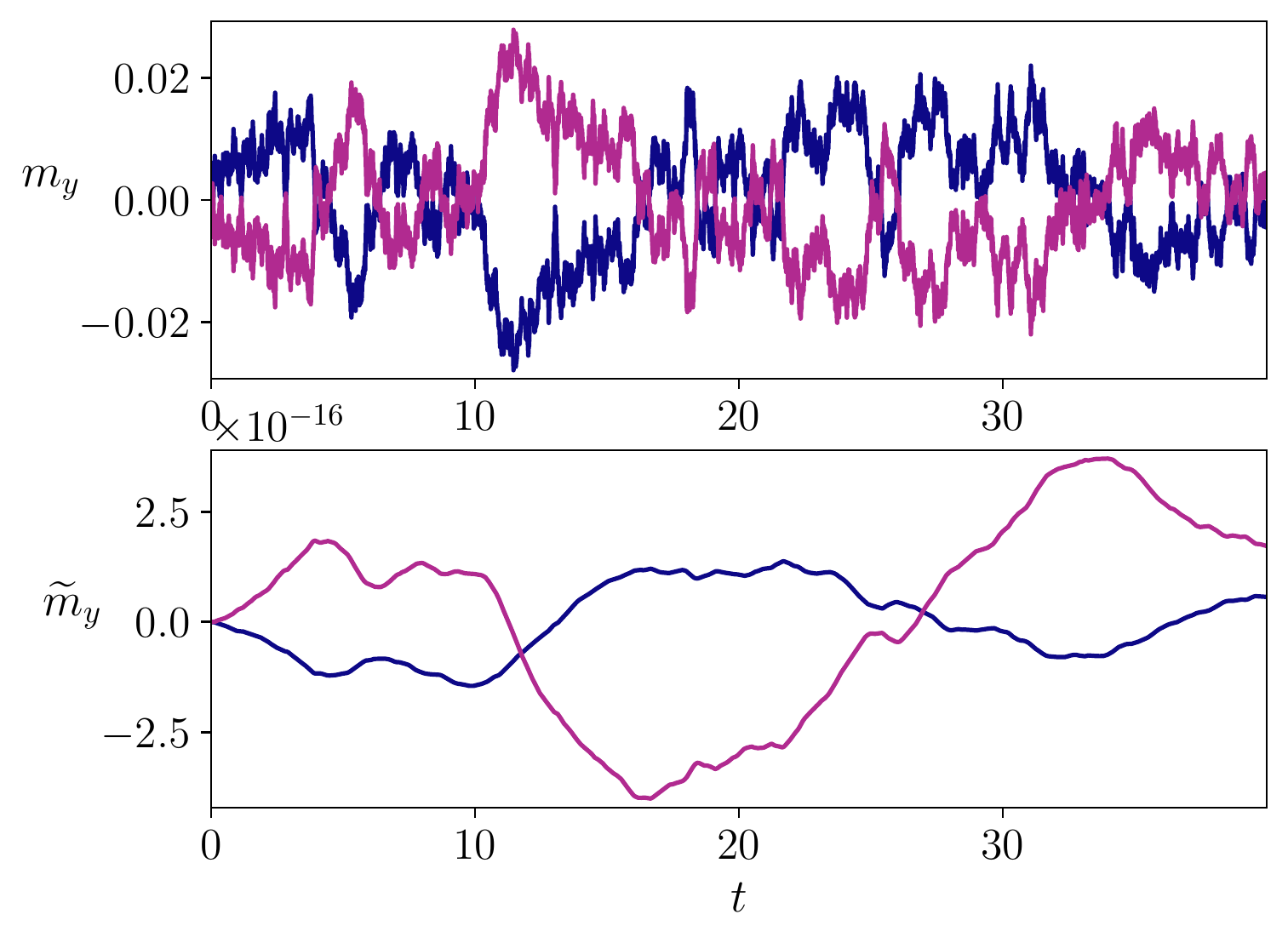} &  \includegraphics[width=6.4cm]{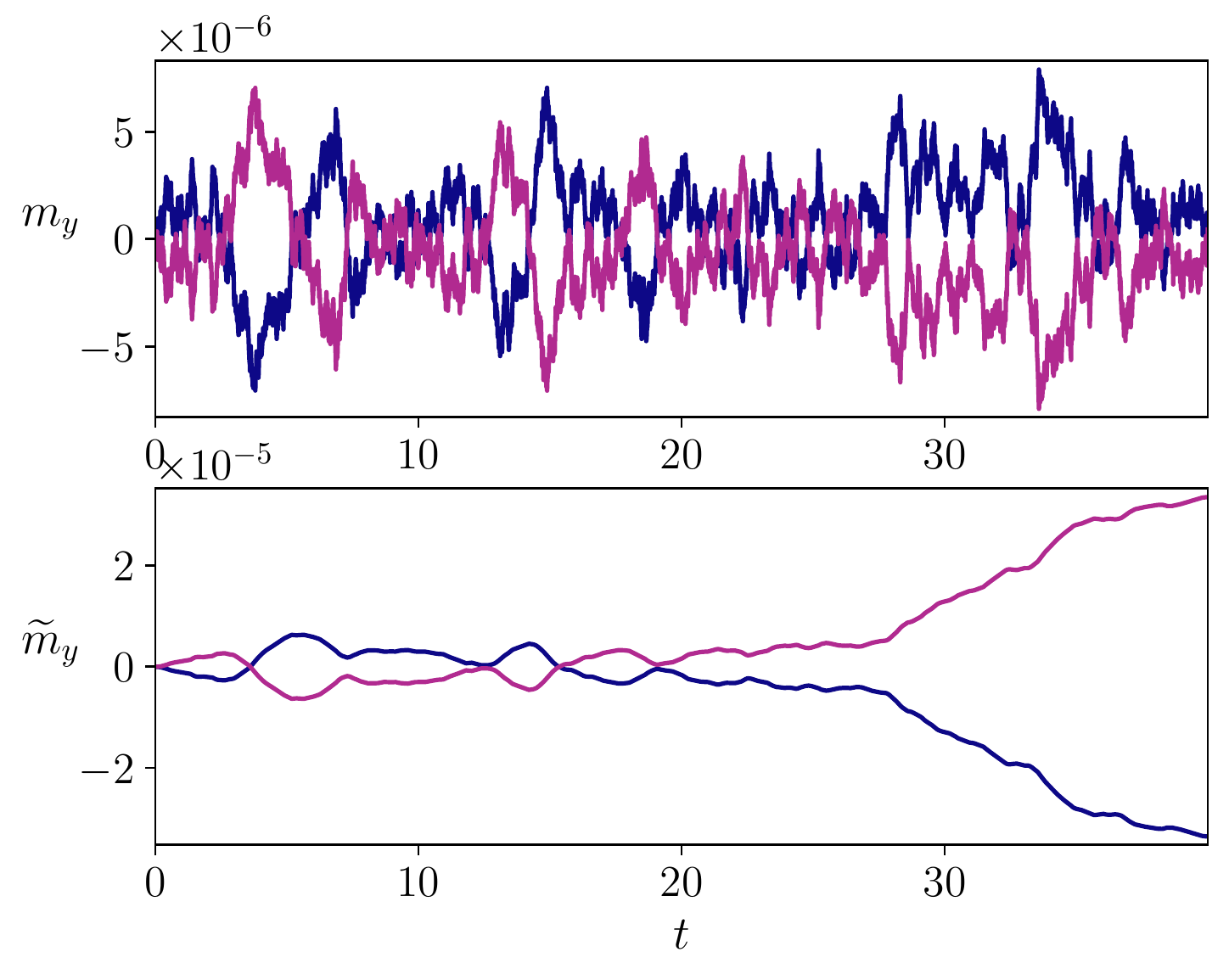}
\end{tabular}
\end{center}
\caption{\textbf{Divergence between system evolution and the marginal flow}. 
Comparison between the true evolution of the system $\*m_y(\*b_t)$ versus the marginal flow $\*{\widetilde m}_y(\*b_t)$ for (\textbf{A})  $\varsigma=0.1$ and random couplings $C_{ij}=\pm 0.1$ with the connection scheme from Fig.~\ref{fig:sm-loops}.C and (\textbf{B})   $\varsigma=10^{-3}$ and random couplings $C_{ij}=\pm 0.1$ with the connection scheme from Fig.~\ref{fig:sm-loops}.B but with $\*C_{yb}=\*0$. In both cases $n_y=n_x =2, n_a=n_s=1$. Note that different parameters (e.g. reducing the noise) yield qualitatively similar results.
} 
 \label{fig:marginal-flow}
 \end{figure}

Furthermore, we can show that this equivalence is also incorrect even in the case of weak couplings. Specifically, for the first order weak coupling approximation with $\*J = -\*I + \*C$ and  $\^\Gamma=\varsigma^{2}\*I$. Assuming $\^{\Sigma}^*_{bb}$ is nonsingular (i.e. $\pr{\^{\Sigma}^*_{bb}}^{g} =\pr{\^{\Sigma}^*_{bb}}^{-1} $), the weak coupling expansion of its inverse according to the Neumann series (other expansions exist in the case of generalized inverses \cite{climent2001geometrical}) is,
\begin{align}
    \pr{\^{\Sigma}^*_{bb}}^{-1} =& \varsigma^{-2} (\*I - \frac{1}{2}(\*C_{bb} + \*C_{bb}^\intercal)) + \mathcal{O}(\*C^2).
\end{align}
Using the weak-coupling  approximation of $\^{\Sigma}^*$ (Eq.~\ref{eq:covariance-expansion}),
Eq.~\ref{eq:most-likely-state-evolution} results in
\begin{align}
     \frac{\mathrm{d}\*m_y(\*b_t)}{\mathrm{d}t} =&  \frac{1}{2}(\*C_{yb} + \*C_{by}^\intercal) (\*I - \frac{1}{2}(\*C_{bb}+\*C_{bb}^\intercal))\big( (-\*I_{bz} + \*C_{bz})(\*z_t-\^\rho) \nonumber\\&+  \^\omega_{b,t}\big)+ \mathcal{O}(\*C^2)
     \nonumber\\ =&  - \frac{1}{2}(\*C_{yb} + \*C_{by}^\intercal) (\*b_t-\^\rho_b - \^\omega_{b,t})+ \mathcal{O}(\*C^2).
\end{align}

In contrast, the marginal flow in Eq.~\ref{eq:marginal-flow-approximation} results in
  \begin{align}
     \frac{\mathrm{d}\*{\widetilde m_y}(\*b_t)}{\mathrm{d}t} =& 
     \Big((-\*I_{yy} + \*C_{yy})\frac{1}{2}(\*C_{yb} + \*C_{by}^\intercal) (\*I - \frac{1}{2}(\*C_{bb}+\*C_{bb}^\intercal))
     \nonumber\\&+ \*C_{yb}\Big)(\*b_t-\^\rho_b )  +  \mathcal{O}(\*C^2)
     \nonumber\\ =& - \frac{1}{2}(-\*C_{yb} + \*C_{by}^\intercal) (\*b_t - \^\rho_b  ) +  \mathcal{O}(\*C^2),
 \end{align}
 which not only ignores the random fluctuations term, but also reverses the sign of the term $\*C_{yb}$. Note that $\*C_{yb}=\*C_{by}^\intercal$ when we force solenoidal uncoupling (Assumption \ref{ass:block-diagonal-Q}, see Section \ref{sec:requirements}), making $\frac{\mathrm{d}\*{\widetilde m_y}(\*b_t)}{\mathrm{d}t} = \*0$ in that case (while $\frac{\mathrm{d}\*m_y(\*b_t)}{\mathrm{d}t}$ is non-zero).
 
 For this weak coupling approximation, the only case in which the approximation is valid is one in which $\varsigma=0$ (a  deterministic system) and $\*C_{yb}=\*0$ (a system where the agent just observes the environment without affecting it).  Similar expressions could be derived for higher order approximations.
 
 As an example of the dissimilarity of these quantities, in Fig.~\ref{fig:marginal-flow}.A we display $\*m_y(\*b_t)$ and $\*{\widetilde m}_y(\*b_t)$ for arbitrary parameters structured as the sensorimotor loop described in Fig.~\ref{fig:sm-loops}.C with random couplings $C_{ij}=\pm 0.1$ (note that some weights are set to zero and others forced to be symmetric) and  $n_y=n_x =2, n_a=n_s=1, \varsigma=0.1$. We see that  $\*{\widetilde m}_y(\*b_t)$ cannot capture the structure in  $\*m_y(\*b_t)$. The reason behind this is that the derivative of $\*{\widetilde m_y}(\*b_t)$ accumulates an error from fluctuations in the system that are not captured, displaying a random walk behaviour that is absent in the real variable $\*m_y(\*b_t)$ .
Similarly, in the most favourable case, setting a very small noise $\varsigma = 10^{-3}$ and $\*C_{yb}=\*0$ (Figure \ref{fig:marginal-flow}.B), the situation is similar, as even very small noise terms are sufficient for driving the two terms apart due to the integration of random fluctuations over time.
 
\subsubsection*{Assumption \ref{ass:bayesian-mechanicsII}: A way out? Problems with interpreting behaviour as Bayesian inference only `on average'}

The results of the previous section suggest that the FEP could be, in practice, inapplicable for describing or explaining the behaviour of living systems.
Some of the most recent works on the FEP \citep{friston2021some} try to avoid the problems described above and state that the free energy principle describes just marginal flows, that is, it describes the behaviour of a system just on average over different trajectories. This could appear to circumvent some issues presented in the previous section, as it implies substituting Assumption \ref{ass:bayesian-mechanics} by a more relaxed interpretation of a gradient descent on the free energy described by Assumption \ref{ass:bayesian-mechanicsII}. 
However, under a close inspection, we encounter that the situation is not improved by this claim.
Assumption \ref{ass:bayesian-mechanicsII} entails two problems: 1) the mapping described by Assumption \ref{ass:sigma-mapping} no longer connects internal and external flows, and 2) the conditional average flow following the gradient of the free energy does not guarantee a gradient descent in the effective behaviour. While the first problem can be solved, the second has deep implications as in most cases the average flow does not describe the true behaviour of a system in the presence of stochastic fluctuations, therefore threatening to render the FEP inapplicable to describe most living systems.
In this last section, we briefly explore the validity of the theory when applied only to the average flows of a system (and not the evolution of the most likely states).

A first, practical problem implied by this claim is that if the FEP only applies to marginal flows $\ang{f_y(\*y_t,\*b_t)  }_{\*b_t}, \ang{f_x(\*b_t,\*x_t)  }_{\*b_t}$, then a new mapping between flows is required.
As we described in Eq.~\ref{eq:derivative-mapping}, a mapping between the dynamics of the most likely states is derived from the chain rule as the gradient of the mapping $\sigma$. 
If instead one interprets the FEP as connecting the conditional average flow of external and internal states a new mapping is required:
\begin{align}
    \ang{f_y(\*y_t,\*b_t)  }_{\*b_t}= \phi(\ang{f_x(\*b_t,\*x_t)  }_{\*b_t} ).
\end{align}
In general, this mapping can take complicated forms, and often a unique mapping will not exist. In linear systems, however, we can simplify the marginal flows
\begin{align}
    \ang{f_y(\*y_t,\*b_t)  }_{\*b_t} =& \*J_{yy} (\*m_y(\*b_t)-\^\rho_y) + \*J_{yb} (\*b_t-\^\rho_b)
    \nonumber\\=& 
    \pr{ \*J_{yy} + (\^\Sigma_{yb}^* (\^\Sigma_{bb}^* )^g)^{-1}} (\*m_y(\*b_t)-\^\rho_y)
    \nonumber\\ =& \pr{ \*J_{yy} + (\^\Sigma_{yb}^* (\^\Sigma_{bb}^* )^g)^{-1}} \^{\Sigma}^*_{yb}\pr{\^{\Sigma}^*_{xb}}^{-1} \pr{\*m_{x}(\*b_t) -\^\rho_x},
    \\ \ang{f_x(\*b_t,\*x_t)  }_{\*b_t}  =& \*J_{xx} (\*m_x(\*b_t)-\^\rho_x) + \*J_{xb} (\*b_t-\^\rho_b) 
    \nonumber\\=& \pr{ \*J_{xx} + (\^\Sigma_{xb}^* (\^\Sigma_{bb}^* )^g)^{-1}} (\*m_x(\*b_t)-\^\rho_x).
\end{align}
Which, if $\^\Sigma_{bb}^*$ is invertible, yields
\begin{align}
    \ang{f_y(\*y_t,\*b_t)  }_{\*b_t}  =& \pr{ \*J_{yy} + \^\Sigma_{bb}^*(\^\Sigma_{yb}^*)^{-1}} \^{\Sigma}^*_{yb}\pr{\^{\Sigma}^*_{xb}}^{-1} 
    \nonumber\\&\cdot \pr{ \*J_{xx} +\^\Sigma_{bb}^* (\^\Sigma_{xb}^* )^{-1}}^{-1}\ang{f_x(\*b_t,\*x_t)  }_{\*b_t} 
    \nonumber\\ \equiv & \phi(\ang{f_x(\*b_t,\*x_t)  }_{\*b_t} ),
\end{align}
that results in the mapping $\phi$ being very different from $\nabla_{\*m_x}\sigma$. Thus, in general,
\begin{align}
   \ang{f_y(\*y_t,\*b_t)  }_{\*b_t}\neq \nabla_{\*m_x}\sigma  \ang{f_x(\*b_t,\*x_t)  }_{\*b_t},
\end{align}
thus contradicting \cite{friston2019free, parr2020markov}. Thus, an interpretation of the FEP over conditional average flows should replace Assumption \ref{ass:sigma-mapping} with a new mapping.
This result, in combination with a block-diagonal Q (Assumption \ref{ass:block-diagonal-Q}) allows rewriting $\ang{ f_y(\*y_t,\*b_t)  }_{\*b_t} $ and $ \ang{f_x(\*x_t,\*b_t)  }_{\*b_t} $ pointing in the direction of free energy minimization, mediated by a mapping $\phi$.

However, there is an important conceptual problem that remains even if a new mapping is derived. 
The FEP relies on finding a variable that behaves as a gradient descent on the free energy functional as described in Eq.~\ref{eq:gradient-descent}.
If Assumption \ref{ass:bayesian-mechanics} is not met, then there is no variable in the system that behaves following a gradient descent on the free energy  (i.e. a variable $\^\theta_t$ with behaviour determined by $ \frac{\mathrm{d}\^\theta_t}{\mathrm{d}t} = -\^\gamma \nabla_{\^\theta} F(\^\theta_t,\*b_t)$).
In this case, this assumption could be relaxed to Assumption \ref{ass:bayesian-mechanicsII} but, as we observed in the results and simulations above, even when the average flow of the system is related with the free energy (i.e. $\ang{ \frac{\mathrm{d}\^\theta_t}{\mathrm{d}t} }_{\*b_t} = -\^\gamma \nabla_{\^\theta} F(\^\theta_t,\*b_t)$) this is not a good description of the behaviour of a system.

In this case, the claim of the FEP, if all other assumptions are hold, can be described by Eq.~\ref{eq:conditional-average-flow-ii}, which relaxes the requirement of a strict gradient descent (described by Eq.~\ref{eq:bayesian-mechanics}) and requires instead that the conditional average flow is directed in the direction of the gradient of the free energy. 
The problem with this relaxed gradient descent interpretation is that it does not guarantee that a system effectively performs a gradient descent, not even \emph{on average}.
In \ref{app:marginal-flows} we illustrate this issue in a simple bivariate linear stochastic model with variables $y,b$. We observe that this particular model presents a global attractor located at $y=b=0$ with solenoidal flows in the form of a spiral flow (Fig.~\ref{fig:2D-flows}.A). In contrast, the conditional average flow suggests a monotonic gradient ascent on $m_y^2(b)$, dismissing solenoidal flows in the system and transforming an attracting flow into a repelling one. This is a simple example showing how, in general, the conditional average flows do not describe the behaviour of the system. 
Moreover, conditional average flows can be misleading and not even a good approximation about the average behaviour of a system, indicating a gradient ascent/descent on some quantity when the behaviour of the system performs the opposite action.
This is also exemplified in recent work simulating linear systems, where the free energy gradient only captures attracting tendencies at highly surprising states, not capturing solenoidal flows nor behaviour near the NESS global attractor \cite{dacosta2021bayesian, parr2021memory}.

To summarize, our results show that, in general, even if a relaxed version of the assumptions of the FEP holds, a system cannot be interpreted \emph{as if} performing Bayesian inference over external states. The reason behind this is that the average flows $\ang{f_y(\*y_t,\*b_t)  }_{\*b_t}$ and $\ang{f_x(\*b_t,\*x_t)  }_{\*b_t}$ do not describe the behaviour of the system, as it can be easily seen from the results in Fig.~\ref{fig:marginal-flow}.
Intuitively, substituting the true flow -- $f_y(\*y_t,\*b_t)$ -- by an average flow fixing the blanket state -- $\ang{f_y(\*y_t,\*b_t)  }_{\*b_t}$  --decouples the trajectory of $\*y$ from its previous state, which in most dynamical systems will result in an impoverished description, not capturing its real, history-dependent, behaviour.



\section{Conclusion}

The latest formulation of the free energy principle \cite{friston2019free, parr2020markov} states that, in any dynamical system equipped with a Markov blanket, the flow of internal states can be construed as a gradient ascent on Bayesian model evidence.
This assumption rests on two crucial moves. The first move connects the system's average flow with a gradient on a variational free energy. This connection relies on the existence of a Markov blanket and the emergence of a particular statistical structure precluding solenoidal couplings.
The second is the interpretation that this relation between free energy and an average flow results in systems behaving \emph{as if} performing variational inference over the states of its environment.
In this review, we have summarized crucial steps required for this claim (Conditions \ref{cond:canonical-flow}--\ref{cond:markov-blanket} and Assumptions \ref{ass:block-diagonal-Q}--\ref{ass:bayesian-mechanics}) and have shown that several of these conditions cannot be met in general by linear, weakly-coupled stochastic systems.

The first step compels a discussion about the generality of the FEP. That is, if the principle requires a particular statistical structure, how general is this structure, and how broadly can we expect it to be present among the class of systems capturing the properties of living and cognitive processes? 
We discover that, in the class of linear systems explored, the answer to this question is that the statistical structure required by the FEP only arises in a very narrow class of systems, requiring stringent conditions such as fully symmetric agent-environment interactions that we cannot, in general, expect from living systems \citep{rothman1977membrane, fadeel2009ins, barandiaran2009defining, ruiz2004basic}.
The generality of the FEP has been questioned in the past due to conceptual issues \cite{raja2021markov, bruineberg2020emperor} or the existence of counterexamples challenging the idea that perception-action interfaces, Markov blankets and solenoidal decoupling follow from each other \cite{biehl2021technical}. However, to our knowledge, our study is the first that shows that the assumptions of the FEP do not hold for a vast class of systems, namely, linear, weakly coupled systems, except for the limited case of fully symmetric agent-environment interaction.

This is concerning for two reasons. First, the FEP is designed for Gaussian (i.e. in most cases linear) stochastic systems \cite{parr2020markov, biehl2021technical}. Thus, our results would imply that, as currently defined, the FEP cannot be fully implemented in a broad set of systems from the class it was designed for.
Second, one could hope that the introduction of strong couplings or non-linearity allows some specific systems to meet the required conditions. Some recent approximations of the behaviour of chaotic oscillators point in this direction \cite{friston2021stochastic}.
However, this claim should be regarded with some scepticism, as the introduction of stronger or higher-order couplings would result in additional terms in the expansions explored in this article (see \ref{app:linear-langeving-dynamics}). For most parameter settings, systems with stronger couplings will not result in the independence relations required for Condition \ref{cond:markov-blanket} and Assumption \ref{ass:block-diagonal-Q}. This will lead, in general, to a more significant divergence between the evolution of the system and its average flow, making it more unlikely that Assumptions \ref{ass:bayesian-mechanics} or \ref{ass:bayesian-mechanicsII} will hold.
We leave it to future work to explore the accuracy of this claim and investigate whether the assumptions of the FEP can be met beyond the class of systems explored here, e.g. displaying strong or nonlinear couplings.

The second step concerns how informative the FEP is about the behaviour of an agent. The FEP justifies that any system can be described as if performing variational inference through the existence of a conditional synchronization manifold relating the direction of the free energy gradient and the evolution of the most likely states of a system.
We observe that this manifold can exist in a broad class of systems. Nevertheless, assuming a strict gradient descent interpretation (Assumption \ref{ass:bayesian-mechanics}), it is problematic to  connect the evolution of the most likely states to the average flow of a system (and, in consequence, to assume a system will behave as if minimizing the variational free energy).
The problem lies in implicitly relating the rate of change of the average (expected) state as being described by the average  flow (the expectation of the rate of change) conditioned on a blanket state. 
If, instead, we consider a more relaxed interpretation of the free energy gradient descent (i.e. just taking place on average, Assumption \ref{ass:bayesian-mechanicsII}), we encounter that new problems arise. First, a new mapping between the flows of internal and external states is required. Second, we observe that the average flow  cannot, in general, describe the true behaviour of a system.
In sum, the FEP as it stands does not do justice to the influence of the system's trajectory in determining its future behaviour. The reason behind this is that the gradient of the free energy defined by the principle is computed for the average of an ensemble of trajectories.
Thus, even when the free energy gradient can be connected with an average flow (which, as we have shown, happens under very specific conditions), this is mainly uninformative about the behaviour of a system subject to stochastic interactions.
This is especially relevant for emergent discussions about the compatibility  of the FEP with enactive and autopoietic theories of cognition (e.g. \cite{bruineberg2018anticipating, constant2021representation, ramstead2021multiscale}). Specifically, enactive principles stress the history-dependence of living systems, and this supposes a fundamental incompatibility with the assumptions of the FEP \cite{dipaolo2021laying}. In particular, enactive views of cognition conceive sense-making as a process emerging from the history of interactions of a system, which is invisible for a gradient of the free energy described as an average flow.

 
 

The motivation behind the FEP aims to connect ideas from variational inference with the dynamics of complex, self-organizing systems. This claim is exceptionally appealing, as it could potentially allow applying the machinery from Bayesian and information theoretical approaches to describe many systems that are intractable in practice. 
However, by inspecting the theory and its assumptions in the context of a broad class of analytically tractable models, we discover that many of the steps required to derive the theory do not straightforwardly follow or present significant conceptual problems that need to be resolved. This finding illustrates the difficulties in developing a theory of life and cognition over interdependent sets of mathematical assumptions, and how testing these assumptions and their relations against tractable models can help overcome these difficulties.


\section*{Acknowledgements}
We are grateful to Karl Friston, Lancelot Da Costa, Thomas Parr, Iñigo Arandia-Romero and Hideaki Shimazaki for their constructive feedback and comments on this manuscript.
M.A. is thankful to Martin Biehl for helpful discussions about references \cite{friston2019free, parr2020markov, friston2021some}. We are also grateful to Manuel Baltieri for his \href{https://pubpeer.com/publications/4D0EB9FF81FF278AC6E2DE78D6E9EE\#1}{open peer-review} of a preprint of this manuscript.

M.A. was funded by the European Union's Horizon 2020 research and innovation programme under a Marie Skłodowska-Curie Action (grant agreement \mbox{892715}) and supported in part by the Basque Government project \mbox{IT 1228-19} and the Spanish Ministry of Science and Innovation project PID2019-104576GB-I00. C.L.B. is supported by BBRSC grant BB/P022197/1.

\bibliographystyle{elsarticle-num} 
\bibliography{refs}

\begin{thebibliography}{10}
\expandafter\ifx\csname url\endcsname\relax
  \def\url#1{\texttt{#1}}\fi
\expandafter\ifx\csname urlprefix\endcsname\relax\def\urlprefix{URL }\fi
\expandafter\ifx\csname href\endcsname\relax
  \def\href#1#2{#2} \def\path#1{#1}\fi

\bibitem{friston2012free}
K.~Friston, P.~Ao, Free energy, value, and attractors, Computational and
  mathematical methods in medicine 2012 (2012).

\bibitem{friston2013life}
K.~Friston, Life as we know it, Journal of the Royal Society Interface 10~(86)
  (2013) 20130475.

\bibitem{friston2019free}
K.~Friston, A free energy principle for a particular physics, arXiv preprint
  arXiv:1906.10184 (2019).

\bibitem{parr2020markov}
T.~Parr, L.~Da~Costa, K.~Friston, Markov blankets, information geometry and
  stochastic thermodynamics, Philosophical Transactions of the Royal Society A
  378~(2164) (2020) 20190159.

\bibitem{hohwy2016self}
J.~Hohwy, The self-evidencing brain, Nous 50~(2) (2016) 259--285.

\bibitem{clark2015surfing}
A.~Clark, Surfing uncertainty: Prediction, action, and the embodied mind,
  Oxford University Press, 2015.

\bibitem{friston2009free}
K.~Friston, The free-energy principle: a rough guide to the brain?, Trends in
  cognitive sciences 13~(7) (2009) 293--301.

\bibitem{friston2010action}
K.~Friston, J.~Daunizeau, J.~Kilner, S.~J. Kiebel, Action and behavior: a
  free-energy formulation, Biological cybernetics 102~(3) (2010) 227--260.

\bibitem{friston2010free}
K.~Friston, The free-energy principle: a unified brain theory?, Nature reviews
  neuroscience 11~(2) (2010) 127--138.

\bibitem{friston2007variational}
K.~Friston, J.~Mattout, N.~Trujillo-Barreto, J.~Ashburner, W.~Penny,
  Variational free energy and the laplace approximation, Neuroimage 34~(1)
  (2007) 220--234.

\bibitem{friston2005theory}
K.~Friston, A theory of cortical responses, Philosophical transactions of the
  Royal Society B: Biological sciences 360~(1456) (2005) 815--836.

\bibitem{buckley2017free}
C.~L. Buckley, C.~S. Kim, S.~McGregor, A.~K. Seth, The free energy principle
  for action and perception: A mathematical review, Journal of Mathematical
  Psychology 81 (2017) 55--79.

\bibitem{millidge2021predictive}
B.~Millidge, A.~Seth, C.~L. Buckley, Predictive coding: a theoretical and
  experimental review, arXiv preprint arXiv:2107.12979 (2021).

\bibitem{friston2015active}
K.~Friston, F.~Rigoli, D.~Ognibene, C.~Mathys, T.~Fitzgerald, G.~Pezzulo,
  Active inference and epistemic value, Cognitive neuroscience 6~(4) (2015)
  187--214.

\bibitem{friston2009reinforcement}
K.~Friston, J.~Daunizeau, S.~J. Kiebel, Reinforcement learning or active
  inference?, PloS one 4~(7) (2009) e6421.

\bibitem{friston2017active}
K.~Friston, T.~FitzGerald, F.~Rigoli, P.~Schwartenbeck, G.~Pezzulo, Active
  inference: a process theory, Neural computation 29~(1) (2017) 1--49.

\bibitem{cullen2018active}
M.~Cullen, B.~Davey, K.~Friston, R.~J. Moran, Active inference in openai gym: A
  paradigm for computational investigations into psychiatric illness,
  Biological psychiatry: cognitive neuroscience and neuroimaging 3~(9) (2018)
  809--818.

\bibitem{millidge2020predictive}
B.~Millidge, A.~Tschantz, C.~L. Buckley, Predictive coding approximates
  backprop along arbitrary computation graphs, arXiv preprint arXiv:2006.04182
  (2020).

\bibitem{hohwy2008predictive}
J.~Hohwy, A.~Roepstorff, K.~Friston, Predictive coding explains binocular
  rivalry: An epistemological review, Cognition 108~(3) (2008) 687--701.

\bibitem{kanai2015cerebral}
R.~Kanai, Y.~Komura, S.~Shipp, K.~Friston, Cerebral hierarchies: predictive
  processing, precision and the pulvinar, Philosophical Transactions of the
  Royal Society B: Biological Sciences 370~(1668) (2015) 20140169.

\bibitem{tschantz2020reinforcement}
A.~Tschantz, B.~Millidge, A.~K. Seth, C.~L. Buckley, Reinforcement learning
  through active inference, arXiv preprint arXiv:2002.12636 (2020).

\bibitem{da2020active}
L.~Da~Costa, T.~Parr, N.~Sajid, S.~Veselic, V.~Neacsu, K.~Friston, Active
  inference on discrete state-spaces: a synthesis, Journal of Mathematical
  Psychology 99 (2020) 102447.

\bibitem{millidge2020deep}
B.~Millidge, Deep active inference as variational policy gradients, Journal of
  Mathematical Psychology 96 (2020) 102348.

\bibitem{fitzgerald2015active}
T.~H. FitzGerald, P.~Schwartenbeck, M.~Moutoussis, R.~J. Dolan, K.~Friston,
  Active inference, evidence accumulation, and the urn task, Neural computation
  27~(2) (2015) 306--328.

\bibitem{parr2019neuronal}
T.~Parr, D.~Markovic, S.~J. Kiebel, K.~Friston, Neuronal message passing using
  mean-field, bethe, and marginal approximations, Scientific reports 9~(1)
  (2019) 1--18.

\bibitem{kappel2021synapse}
D.~Kappel, C.~Tetzlaff, A synapse-centric account of the free energy principle,
  arXiv preprint arXiv:2103.12649 (2021).

\bibitem{tschantz2020learning}
A.~Tschantz, A.~K. Seth, C.~L. Buckley, Learning action-oriented models through
  active inference, PLoS computational biology 16~(4) (2020) e1007805.

\bibitem{calvo2017predicting}
P.~Calvo, K.~Friston, Predicting green: really radical (plant) predictive
  processing, Journal of the Royal Society Interface 14~(131) (2017) 20170096.

\bibitem{friston2006free}
K.~Friston, J.~Kilner, L.~Harrison, A free energy principle for the brain,
  Journal of physiology-Paris 100~(1-3) (2006) 70--87.

\bibitem{friston2021some}
K.~Friston, L.~Da~Costa, T.~Parr, Some interesting observations on the free
  energy principle, Entropy 23~(8) (2021) 1076.

\bibitem{pearl1988probabilistic}
J.~Pearl, Probabilistic reasoning in intelligent systems: networks of plausible
  inference, Elsevier, 1988.

\bibitem{richardson1996automated}
T.~S. Richardson, P.~Spirtes, et~al., Automated discovery of linear feedback
  models, Carnegie Mellon, 1996.

\bibitem{bruineberg2020emperor}
J.~Bruineberg, K.~Dolega, J.~Dewhurst, M.~Baltieri, The emperor's new markov
  blankets, Behavioral and Brain Sciences (2021) 1–63.

\bibitem{biehl2021technical}
M.~Biehl, F.~A. Pollock, R.~Kanai, A technical critique of some parts of the
  free energy principle, Entropy 23~(3) (2021) 293.

\bibitem{rothman1977membrane}
J.~E. Rothman, J.~Lenard, Membrane asymmetry, Science 195~(4280) (1977)
  743--753.

\bibitem{buzsaki2006rhythms}
G.~Buzsaki, Rhythms of the Brain, Oxford University Press, 2006.

\bibitem{fadeel2009ins}
B.~Fadeel, D.~Xue, The ins and outs of phospholipid asymmetry in the plasma
  membrane: roles in health and disease, Critical reviews in biochemistry and
  molecular biology 44~(5) (2009) 264--277.

\bibitem{barandiaran2009defining}
X.~E. Barandiaran, E.~Di~Paolo, M.~Rohde, Defining agency: Individuality,
  normativity, asymmetry, and spatio-temporality in action, Adaptive Behavior
  17~(5) (2009) 367--386.

\bibitem{ruiz2004basic}
K.~Ruiz-Mirazo, A.~Moreno, Basic autonomy as a fundamental step in the
  synthesis of life, Artificial life 10~(3) (2004) 235--259.

\bibitem{kwon2011nonequilibrium}
C.~Kwon, P.~Ao, Nonequilibrium steady state of a stochastic system driven by a
  nonlinear drift force, Physical Review E 84~(6) (2011) 061106.

\bibitem{yuan2017sde}
R.~Yuan, Y.~Tang, P.~Ao, Sde decomposition and a-type stochastic interpretation
  in nonequilibrium processes, Frontiers of Physics 12~(6) (2017) 1--9.

\bibitem{friston2021stochastic}
K.~Friston, C.~Heins, K.~Ueltzh{\"o}ffer, L.~Da~Costa, T.~Parr, Stochastic
  chaos and markov blankets, Entropy 23~(9) (2021) 1220.

\bibitem{parr2021memory}
T.~Parr, L.~Da~Costa, C.~Heins, M.~J.~D. Ramstead, K.~Friston, Memory and
  markov blankets, Entropy 23~(9) (2021) 1105.

\bibitem{reid2015approximate}
N.~Reid, Approximate likelihoods, Proceedings of the ICIAM (2015).

\bibitem{kim2018recognition}
C.~S. Kim, Recognition dynamics in the brain under the free energy principle,
  Neural computation 30~(10) (2018) 2616--2659.

\bibitem{dacosta2021neural}
L.~Da~Costa, T.~Parr, B.~Sengupta, K.~Friston, Neural dynamics under active
  inference: Plausibility and efficiency of information processing, Entropy
  23~(4) (2021) 454.

\bibitem{vatiwutipong2019alternative}
P.~Vatiwutipong, N.~Phewchean, Alternative way to derive the distribution of
  the multivariate ornstein--uhlenbeck process, Advances in Difference
  Equations 2019~(1) (2019) 1--7.

\bibitem{godreche2018characterising}
C.~Godr{\`e}che, J.-M. Luck, Characterising the nonequilibrium stationary
  states of ornstein--uhlenbeck processes, Journal of Physics A: Mathematical
  and Theoretical 52~(3) (2018) 035002.

\bibitem{dacosta2021bayesian}
L.~Da~Costa, K.~Friston, C.~Heins, G.~A. Pavliotis, Bayesian mechanics for
  stationary processes, arXiv preprint arXiv:2106.13830 (2021).

\bibitem{climent2001geometrical}
J.~J. Climent, N.~Thome, Y.~Wei, A geometrical approach on generalized inverses
  by neumann-type series, Linear algebra and its applications 332 (2001)
  533--540.

\bibitem{raja2021markov}
V.~Raja, D.~Valluri, E.~Baggs, A.~Chemero, M.~L. Anderson, The markov blanket
  trick: On the scope of the free energy principle and active inference,
  Physics of Life Reviews (2021).

\bibitem{bruineberg2018anticipating}
J.~Bruineberg, J.~Kiverstein, E.~Rietveld, The anticipating brain is not a
  scientist: the free-energy principle from an ecological-enactive perspective,
  Synthese 195~(6) (2018) 2417--2444.

\bibitem{constant2021representation}
A.~Constant, A.~Clark, K.~J. Friston, Representation wars: Enacting an
  armistice through active inference, Frontiers in Psychology 11 (2021) 3798.

\bibitem{ramstead2021multiscale}
M.~J. Ramstead, M.~D. Kirchhoff, A.~Constant, K.~J. Friston, Multiscale
  integration: beyond internalism and externalism, Synthese 198~(1) (2021)
  41--70.

\bibitem{dipaolo2021laying}
E.~Di~Paolo, E.~Thompson, R.~D. Beer, Laying down a forking path:
  Incompatibilities between enaction and the free energy principle, PsyArXiv
  (2021).

\end{thebibliography}
\newpage

\appendix

\setcounter{figure}{0}    
\section{Mathematical definitions of main concepts of the free energy principle}
\label{app:definitions}

\textbf{Perception-action partition.} The FEP assumes that the system can be decomposed into external, sensory, active and internal states, $\*z=\{\*y,\*s,\*a,\*x\}$, configured as a perception-action loop reflecting an interface mediating between `autonomous' states (active and internal states $\{\*a,\*x\}$) and `non-autonomous' states (external and sensory states $\{\*y,\*s\}$). This leads to describing the evolution of the system as
\begin{align}
\frac{\mathrm{d}\*z_t} {\mathrm{d}t} =& f(\*z_t) + \^\omega_t,
\\ f(\*z_t) =&
\begin{bmatrix}
    f_y(\*y_t,\*s_t,\*a_t) \\
    f_s(\*y_t,\*s_t,\*a_t) \\
    f_a(\*s_t,\*a_t,\*x_t) \\
    f_x(\*s_t,\*a_t,\*x_t) 
\end{bmatrix}.
\end{align}

\textbf{NESS and solenoidal flows.}
The FEP assumes that the system will reach a non-equilibrium steady state described by the probability density function $p(\*z_t)$, which can be described using  a SDE decomposition that separates the flow into dissipative ($-\^\Gamma \nabla_{\*z} \textfrak{J}(\*z_t)$) and solenoidal ($\*Q\nabla_{\*z} \textfrak{J}(\*z_t)$) components 
    \begin{align}
        f(\*z_t) =& (\*Q - \^\Gamma)\nabla_{\*z} \textfrak{J}(\*z_t),
        \\ \textfrak{J}(\*z_t) =& -\log p(\*z_t),
    \end{align}
The FEP makes some requirements about the solenoidal flow matrix $\*Q$, as it assumes it is state-independent (i.e. $\*Q$ does not change with $\*z$) and that it is sparse in the sense that couplings between some states (e.g. internal and external) are zero (see \ref{ass:block-diagonal-Q}).

\textbf{Markov blanket.}
The steady state distribution of the system is  described in terms of a Markov blanket, where internal/external states are independent when conditioned on its blanket states.
\begin{align}
    p(\*y_t,\*x_t|\*b_t) = p(\*y_t|\*b_t) p(\^x_t|\*b_t).
\end{align}

\textbf{Conditional synchronisation manifold.}
The FEP assumes that there is a smooth and differentiable function, $\sigma$, which maps between the most likely internal and external states given a blanket state,
    \begin{align}
    \*m_y(\*b_t) = \sigma(\*m_x(\*b_t) ),
    \end{align}
and the gradient $\nabla_{\*m_x} \sigma(\*m_x)$ is invertible (i.e. $(\nabla_{\*m_x} \sigma(\*m_x))^{-1}$ exists).
The FEP generally refers to this mapping as a \emph{conditional synchronisation manifold}, and proposes that its existence allows to
characterise the relationship between (maximum a posteriori) internal and external states in terms of internal states `sensing' or `tracking' external states through the Markov blanket \cite{friston2019free}.

\textbf{Rate of change of the average.}
By virtue of the conditional synchronisation manifold, the rate of change of the average internal and external states of the system are connected by the gradients of the mapping function.
\begin{align}
    \*m_y(\*b_t) = \sigma(\*m_x(\*b_t) ) \Rightarrow \frac{\mathrm{d}\*m_y(\*b_t)}{\mathrm{d}t} = \nabla_{\*m_x} \sigma \frac{\mathrm{d}\*m_x(\*b_t)}{\mathrm{d}t} .
\end{align}

\textbf{Variational free energy.}
A system performing Bayesian inference tries to minimize the surprise of observed states ($-\log p(\*b)$), according to an internal model. 
However, a system cannot access this surprise value without complete knowledge of its environment, Bayesian inference prescribes to use a lower bound of this surprise described by the variational free energy $F(\^\theta,\*b)$:
\begin{align}
	-\log p(\*b) < F(\^\theta,\*b) =& -\log p(\*b) + D_{KL}(q(\*y|\^\theta)||p(\*y|\*b)),
\end{align}
which is composed of the surprise plus a term capturing the distance from the probability of external states given the blanket $p(\*y|\*b)$ to an internal model of the world $q(\*y|\^\theta)$ parametrized by  $\^\theta$.

\textbf{Average flow and the free energy gradient.}
The FEP proposes that the evolution of internal and external states of a system can be described, under a Gaussian approximation, as a gradient of the free energy of the system respect to its sufficient statistics.
\begin{align}
    \ang{ f_y(\*y_t,\*b_t)  }_{\*b_t} =&  (\*Q_{yy} - \^\Gamma_{yy}) \ang{\nabla_{\*y}  \textfrak{J}(\*z_t)  }_{\*b_t}
     \nonumber\\ \approx& (\*Q_{yy} - \^\Gamma_{yy}) \nabla_{\*m_y} F(\^\theta,\*b) .
\end{align}
Then, the conditional synchronization manifold is proposed to connect the internal and external average flows (i.e. the gradients $\nabla_{\*m_y}$ and $\nabla_{\*m_x}$), suggesting that internal states are behaving as if trying to perform Bayesian inference over external states.
    
\section{Solution of the linear Langevin dynamics.}
\label{app:linear-langeving-dynamics}
We start with the Ornstein-Uhlenbeck process
\begin{equation}
    \mathrm{d}\*z_t = \*J ( \*z_t -\^\rho)\mathrm{d}t +  \mathrm{d}\*w_t,
\end{equation}
which can be approximated by the equivalent Langevin dynamics
\begin{equation}
    \frac{\mathrm{d}\*z_t}{\mathrm{d}t} = \*J ( \*z_t -\^\rho)+ \^\omega_t,
\end{equation}
where $\*J$ is an $n\times n$ invertible real matrix, $\^\rho$ is an $n$ dimensional real vector, $\*w_t$ is a standard n-dimensional Wiener process, $\^\omega_t$ is a standard n-dimensional Gaussian white noise with covariance matrix $2\^\Gamma$.

The model can be solved using standard methods for systems of differential equations
 \begin{align}
      \mathrm{e}^{-\*J t}   \frac{\mathrm{d} \*z_t }{d t} - \mathrm{e}^{-\*J t}  \*J (\*z_t - \^\rho) =&\mathrm{e}^{-\*J t}   \^\omega_t,
      \\ \frac{\mathrm{d}}{d t}(\mathrm{e}^{-\*J t}(\*z_t - \^\rho)   ) =& \mathrm{e}^{-\*J t}  \^\omega_t ,
      \\ \*z_t  =&  \mathrm{e}^{\*J t} (\*z_0 - \^\rho)+ \^\rho + \int_0^t \mathrm{d}t' \mathrm{e}^{-\*J (t'-t)} \^\omega_{t'}.
 \end{align}

This solution of the system \cite{vatiwutipong2019alternative, godreche2018characterising} yields statistical moments
  \begin{align}
      \*m_t  =& \int d\^\omega p(\^\omega ) \*z_t
      \nonumber\\ =&  \mathrm{e}^{\*J t}\*z_0 + (\*I - \mathrm{e}^{\*J t}) \^\rho,
      \\ \^\Sigma_{t} =& \int d\^\omega p(\^\omega ) ( \*z_t - \*m_t)( \*z_t - \*m_t)^\intercal
      \nonumber\\  =& \int d\^\omega p(\^\omega )\int_0^t \mathrm{d}t'\int_0^t \mathrm{d}t''    \mathrm{e}^{\*J(t-t')} \^\omega_{t'} \^\omega^\intercal_{t''} \mathrm{e}^{\*J^\intercal  (t-t'')} 
      \nonumber\\ =& 2 \int_0^t \mathrm{d}t'\mathrm{e}^{-\*J (t'-t)} \^\Gamma \mathrm{e}^{-\*J^\intercal (t'-t)} .
 \end{align}

These equations are hard to solve analytically. However, we can find the differential equations that result in equivalent integrals, obtaining the time evolution of statistical moments
\begin{align}
      \frac{\mathrm{d} \*m_t }{d t} =&   \mathrm{e}^{\*J t} \*J (\*z_0- \^\rho)
           = \*J (\*m_t - \^\rho ),
      \\ \frac{\mathrm{d} \^\Sigma_{zz,t}}{d t} =&  2 \^\Gamma
       + \^\Sigma_{zz,t} \*J^\intercal +   \*J\^\Sigma_{zz,t} .
 \end{align}
 
 At equilibrium or at the NESS, when the solution is unique, the system stabilizes to the values that make the derivatives equal to zero
 \begin{align}
     \lim_{t\to\infty} \*m_t =& \^\rho,
     \\ \lim_{t\to\infty} \^\Sigma_{zz,t} =& \^{\Sigma}^*, \qquad  \*J \^{\Sigma}^* +\^{\Sigma}^* \*J^\intercal + 2\^\Gamma = 0.
 \end{align}
 where $\^{\Sigma}^*$ can be found numerically solving the above \emph{continuous Lyapunov} equation (a particular case of a \emph{Sylvester} equation).
If $\*J$ is symmetric, the steady state of the system is a state of equilibrium with $\^{\Sigma}^* = - \*J^{-1}\^\Gamma$. However, the FEP focuses instead on NESS, which are more appropriate for describing living systems.

 \subsection{Weak coupling approximation}

In order to study the solution at the NESS, we assume weak couplings of the form $\*J = -\*I + \*C$, with $\*C^2$ being small. We derive
\begin{align}
      2\^\Gamma =& -\*J \^{\Sigma}^* -\^{\Sigma}^* \*J^\intercal = 2\^{\Sigma}^* - \*C \^{\Sigma}^*  - \^{\Sigma}^*  \*C^\intercal ,
     \\ \^{\Sigma}^*  =& \^\Gamma +\frac{1}{2}( \*C \^{\Sigma}^*  + \^{\Sigma}^*  \*C^\intercal ).
\end{align}
Recursively substituting $\^{\Sigma}^*$ in the right-hand side with $\^\Gamma + \frac{1}{2}\*C \^{\Sigma}^*  + \frac{1}{2}\^{\Sigma}^*  \*C^\intercal $, we obtain the time series expansion
\begin{align}
     \^{\Sigma}^*  =&  \^\Gamma + \frac{1}{2}(\*C\^\Gamma+\^\Gamma\*C^\intercal) + \frac{1}{4}\pr{\*C^2\^\Gamma + 2\*C\^\Gamma \*C^\intercal + \^\Gamma(\*C^\intercal)^2 } 
     \nonumber\\ &+ \frac{1}{8}\pr{\*C^3\^\Gamma + 3\*C^2\^\Gamma \*C^\intercal +  3\*C\^\Gamma (\*C^\intercal)^2 + \^\Gamma(\*C^\intercal)^3 }
      + \mathcal{O}(\*C^4).
\end{align}

In the equilibrium case of symmetric couplings, the Hessian is trivially $\*H = \*J \^\Gamma^{-1}$. For systems at a NESS, for $\*J = -\*I + \*C$ and under a uniform noise $\^\Gamma = \varsigma^{-2}\*I$, the inverse covariance (Hessian) can be computed as a Neumann series,
\begin{align}
    \*H =& \pr{ \^{\Sigma}^*}^{-1} =
    \varsigma^{-2}\sum_{k=0}^\infty  \Big(-\frac{1}{2}(\*C+\*C^\intercal)  - \frac{1}{4}\pr{\*C^2 + 2\*C \*C^\intercal + (\*C^\intercal)^2 }  \nonumber\\ &- \frac{1}{8}\pr{\*C^3 + 3\*C^2 \*C^\intercal +  3\*C (\*C^\intercal)^2 + (\*C^\intercal)^3 } + \mathcal{O}(\*C^4)\Big)^k 
    \nonumber\\ =&\varsigma^{-2} \*I -\frac{\varsigma^{-2}}{2}(\*C+\*C^\intercal)  - \frac{\varsigma^{-2}}{4}\pr{\*C^2 + 2\*C \*C^\intercal + (\*C^\intercal)^2 }
    \nonumber\\ & + \frac{\varsigma^{-2}}{4}\pr{\*C^2 + \*C \*C^\intercal + \*C^\intercal  \*C + (\*C^\intercal)^2 }
    \nonumber\\ & - \frac{\varsigma^{-2}}{8}\pr{\*C^3 + 3\*C^2 \*C^\intercal +  3\*C (\*C^\intercal)^2 + (\*C^\intercal)^3 }
    \nonumber\\ & - \frac{\varsigma^{-2}}{8}\big(
    \*C^3 + \*C^2 \*C^\intercal + \*C \*C^\intercal  \*C + \*C(\*C^\intercal)^2 
    \nonumber\\ &+\*C^\intercal \*C^2  + \*C^\intercal \*C \*C^\intercal + (\*C^\intercal)^2  \*C  + (\*C^\intercal)^3 \Big)
   \nonumber\\ &  + \frac{\varsigma^{-2}}{8}\Big(
   2\*C^3 
   + 3\*C^2 \*C^\intercal  + 2\*C \*C^\intercal  \*C 
    + 3 \*C(\*C^\intercal)^2 \nonumber\\ & + \*C^\intercal \*C^2 
    + 2\*C^\intercal \*C\*C^\intercal  + (\*C^\intercal)^2 \*C
      +2 (\*C^\intercal)^3 \Big)
    + \mathcal{O}(\*C^4)
    \nonumber\\ =& 
    \varsigma^{-2} \*I -\frac{\varsigma^{-2}}{2}(\*C+\*C^\intercal)  + \frac{\varsigma^{-2}}{4}\pr{ \*C^\intercal \*C - \*C \*C^\intercal }  
    \nonumber\\ & + \frac{\varsigma^{-2}}{8}\pr{
     \*C  \pr{ \*C^\intercal \*C - \*C \*C^\intercal }   + \pr{ \*C^\intercal \*C - \*C \*C^\intercal }  \*C^\intercal )} + \mathcal{O}(\*C^4).
     \label{eq-app:hessian-expansion}
\end{align}

The submatrix of the Hessian for $\*y,\*x$ couplings is
\begin{align}
    \*H_{yx} =& - \frac{\varsigma^{-2}}{4}\Big( \*C_{ys}  \*C_{xs}^\intercal +  \*C_{ya}  \*C_{xa}^\intercal \Big)
    \nonumber\\& + \frac{\varsigma^{-2}}{8}\Big(   \*C_{ya}(- \*C_{aa}  \*C_{xa}^\intercal -  \*C_{as}  \*C_{xs}^\intercal -  \*C_{ax}  \*C_{xx}^\intercal +  \*C_{aa}^\intercal  \*C_{ax} -  \*C_{aa}^\intercal  \*C_{xa}^\intercal \nonumber\\&  -  \*C_{sa}^\intercal  \*C_{xs}^\intercal +  \*C_{xa}^\intercal  \*C_{xx} -  \*C_{xa}^\intercal  \*C_{xx}^\intercal) 
   \nonumber\\& + \*C_{ys}(-  \*C_{sa}  \*C_{xa}^\intercal -  \*C_{ss}  \*C_{xs}^\intercal +  \*C_{as}^\intercal  \*C_{ax} -  \*C_{as}^\intercal  \*C_{xa}^\intercal -  \*C_{ss}^\intercal  \*C_{xs}^\intercal +  \*C_{xs}^\intercal  \nonumber\\&  \*C_{xx} -  \*C_{xs}^\intercal  \*C_{xx}^\intercal)
    + \*C_{yy}(-  \*C_{ya}  \*C_{xa}^\intercal -  \*C_{ys}  \*C_{xs}^\intercal - \*C_{sy}^\intercal  \*C_{xs}^\intercal)
  \nonumber\\& + \*C_{sy}^\intercal( \*C_{sa}  \*C_{xa}^\intercal +  \*C_{ss}  \*C_{xs}^\intercal) 
   + \*C_{yy} ^\intercal ( \*C_{ya}  \*C_{xa}^\intercal +  \*C_{ys}  \*C_{xs}^\intercal)\Big)  + \mathcal{O}(\*C^4)
    \label{eq-app:hessian_yx-expansion}
\end{align}

\subsection{Solenoidal flows}
\label{app:solenoidal-flow}

 We can describe the surprise of the system  and its gradient as
\begin{align}
    \textfrak{J}(\*z) =&  -\log  p(\*z)
    \nonumber\\  =& \frac{1}{2} (\*z-\*m_z)^\intercal \^\Sigma_{zz}^{-1}(\*z-\*m_z) + \frac{k}{2}\log\pr{2\pi} + \frac{1}{2}\log |\^\Sigma_{zz}|,
    \\ \nabla_{\*z} \textfrak{J}(\*z) =& \pr{\^{\Sigma}^*}^{-1}(\*z-\*m_z).
\end{align}
where we know from Eq.~\ref{eq:NESS-solution} that $ \^{\Sigma}^* = -2\*J^{-1}\^\Gamma  - \*J^{-1} \^{\Sigma}^*  \*J^\intercal $.

At the NESS, the solution can be described in two terms consistent with a SDE  decomposition:
\begin{align}
    \*J(\*z-\^\rho) =&(\*Q - \^\Gamma)  (-2\*J^{-1}\^\Gamma  - \*J^{-1} \^{\Sigma}^* \*J^\intercal)^{-1}(\*z-\^\rho),
    \\ \*Q =& \^\Gamma - \*J(2\*J^{-1}\^\Gamma  + \*J^{-1} \^{\Sigma}^* \*J^\intercal)
     = -\^{\Sigma}^* \*J^\intercal - \^\Gamma    = \^\Gamma +     \*J \^{\Sigma}^*.
\end{align}

Rearranging the terms in the equation above to substitute $\^{\Sigma}^*= \*J^{-1}(\*Q -  \^\Gamma )$ we obtain the equivalence 
\begin{align}
    \*J \*Q  + \*Q \*J^\intercal = \*J \^\Gamma  - \^\Gamma \*J^\intercal,
\end{align}
which again takes the form of a continuous Lyapunov equation. This equation can be solve numerically, or analytically using a power series expansion of $\*J = -\*I + \*C$, expanding the expression
\begin{align}
    \*Q = \frac{1}{2}(\*C \^\Gamma  - \^\Gamma \*C^\intercal) + \frac{1}{2}(\*C \*Q  + \*Q  \*C^\intercal ),
\end{align}
into the power series
\begin{align}
      \*Q  &=  \frac{1}{2}(\*C \^\Gamma  - \^\Gamma \*C^\intercal) + \frac{1}{4}\pr{\*C^2 \^\Gamma - \^\Gamma (\*C^\intercal)^2 }
      \nonumber\\&+ \frac{1}{8}\pr{\*C^3 \^\Gamma + \*C^2 \^\Gamma\*C^\intercal  - \*C \^\Gamma(\*C^\intercal)^2 - \^\Gamma (\*C^\intercal)^3 } + \mathcal{O}(\*C^4).
\end{align}

\subsection{Mapping between the most likely internal and external states}
\label{app:sigma-mapping}

 If $\*z$ can be divided into blanket states $\*b$, and internal/external states $\*h=\{\*x,\*y\}$, the conditional distribution given blanket states is a Normal distribution with moments
 \begin{align}
     \*m_{h}(\*b_t)  =&   \^\rho_h + \^{\Sigma}^*_{hb} \pr{\^{\Sigma}^*_{bb}}^{g} ( \*b_t -\^\rho_b) ,
     \\  \^\Sigma_{hh}(\*b_t) =& \^{\Sigma}^*_{hh}-\^{\Sigma}^*_{hb}\pr{\^{\Sigma}^*_{bb}}^{g}\^{\Sigma}^*_{bh}.
 \end{align}
 Note that the Markov blanket imposes that $\^\Sigma_{xy}(\*b_t) = \^\Sigma_{yx}(\*b_t) =0$, since $\*x,\*y$ are conditionally independent.
 
 Similarly, we can decompose $\*h$ into an internal and external state
 \begin{align}
     \*m_{x}(\*b_t) =&   \^\rho_x + \^{\Sigma}^*_{xb} \pr{\^{\Sigma}^*_{bb}}^{g} ( \*b_t - \^\rho_b) ,
     \\ \*m_{y}(\*b_t) =&    \^\rho_y + \^{\Sigma}^*_{yb} \pr{\^{\Sigma}^*_{bb}}^{g} ( \*b_t - \^\rho_b).
 \end{align}
 If the mapping from $\*b$ to $\*x$ is injective (for this $n_x\geq n_b$ is required), the first term can be rearranged into 
  \begin{align}
    \*b_t - \^\rho_b = \^{\Sigma}^*_{bb} \pr{\^{\Sigma}^*_{xb}}^{-1} \pr{\*m_{x}(\*b_t) -\^\rho_x},
 \end{align}
 yielding
  \begin{align}
     \*m_{y}(\*b_t) =&  \*m_y + \^{\Sigma}^*_{yb} \pr{\^{\Sigma}^*_{bb}}^{g}\^{\Sigma}^*_{bb} \pr{\^{\Sigma}^*_{xb}}^{-1} \pr{\*m_x(\*b_t) -\^\rho_x},
     \nonumber\\ =&  \^\rho_y + \^{\Sigma}^*_{yb} \pr{\^{\Sigma}^*_{xb}}^{-1} \pr{\*m_x(\*b_t) -\^\rho_x} \equiv \sigma(\*m_x(\*b_t)),
 \end{align}
which is a linear mapping of $\*m_x(\*b_t)$ to $\*m_y(\*b_t)$.

\begin{align}
     \*m_{x}(\*b_t) =&  \*m_x + \^{\Sigma}^*_{xb} \pr{\^{\Sigma}^*_{bb}}^{g}\^{\Sigma}^*_{bb} \pr{\^{\Sigma}^*_{yb}}^{-1} \pr{\*m_{y}(\*b_t) -\*m_y(\*b_t)}
     \nonumber\\ =&  \*m_x + \^{\Sigma}^*_{xb} \pr{\^{\Sigma}^*_{yb}}^{-1} \pr{\*m_{y}(\*b_t) -\*m_y(\*b_t)} \equiv \sigma^{-1}(\*m_y(\*b_t)).
 \end{align}
 
 In the NESS limit
 \begin{align}
     \frac{\mathrm{d}\*m_{y}(\*b_t)}{\mathrm{d}t} =&  \^{\Sigma}^*_{yb} \pr{\^{\Sigma}^*_{xb}}^{-1}  \frac{\mathrm{d}\*m_{x}(\*b_t)}{\mathrm{d}t} = \nabla_{\*m_x} \sigma(\*m_{x}(\*b_t)) \frac{\mathrm{d}\*m_{x}(\*b_t)}{\mathrm{d}t} .
 \end{align}
 Similarly, if the mapping from $\*b$ to $\*y$ is injective (for this $n_y\geq n_b$ is required), we can invert the relation
  \begin{align}
    \frac{\mathrm{d}\*m_{x}(\*b_t)}{\mathrm{d}t} =&  \^{\Sigma}^*_{xb} \pr{\^{\Sigma}^*_{yb}}^{-1}  \frac{\mathrm{d}\*m_{y}(\*b_t)}{\mathrm{d}t} = \nabla_{\*m_y} \sigma^{-1}(\*m_{y}(\*b_t)) \frac{\mathrm{d}\*m_{y}(\*b_t)}{\mathrm{d}t}.
 \end{align}

\section{How informative are conditional marginal flows? A minimal example}
\label{app:marginal-flows}
\begin{figure}
\begin{center}
     \includegraphics[width=10cm]{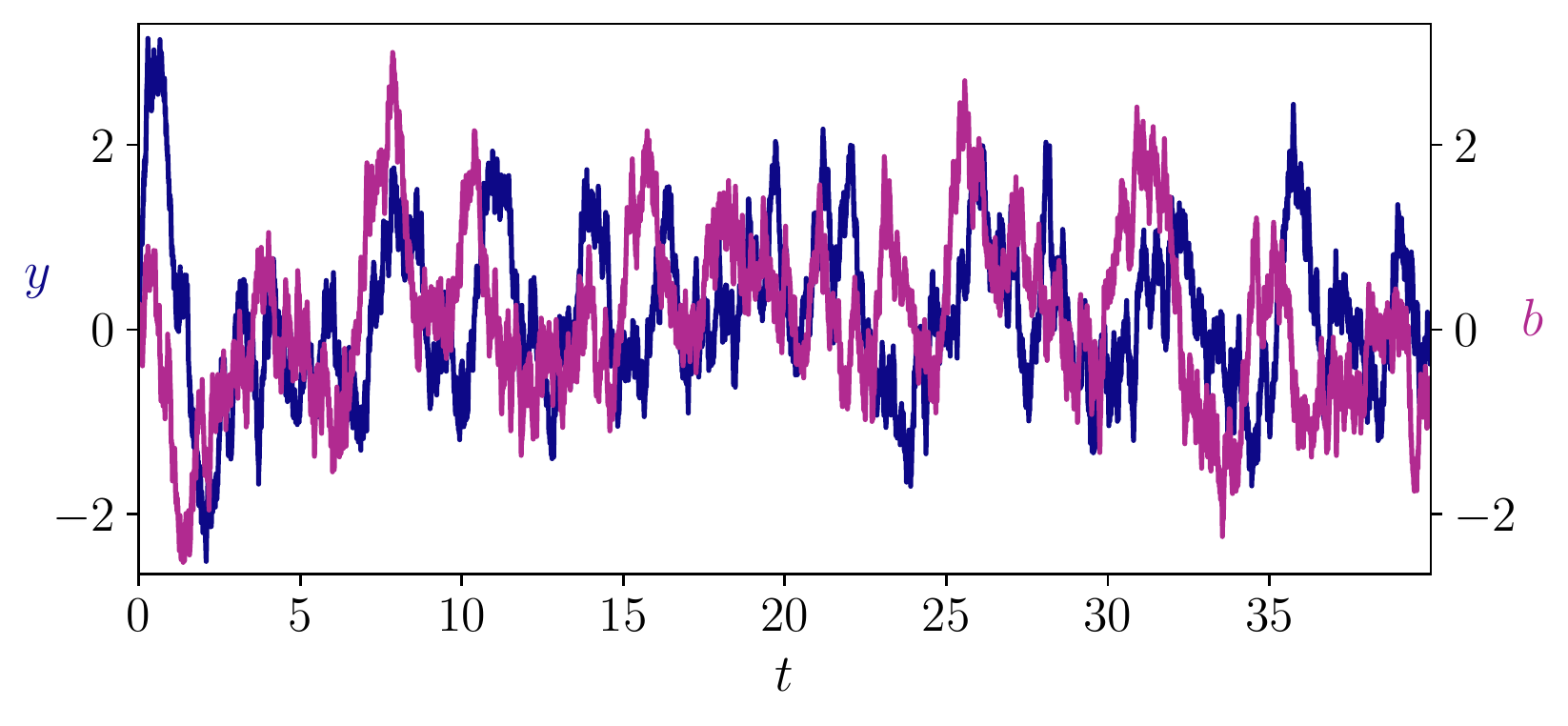} 
\end{center}
\caption{\textbf{Evolution of a bivariate linear stochastic system}. 
Example trajectory of the system described by Eq.~\ref{eq-app:langevin-dynamics} with parameters $\varsigma=1$, $J_{yy}=J_{bb}=-1$, $J_{yb}= 1.2 $, $J_{by}= -1 $.
} 
 \label{fig:2D-timeseries}
 \end{figure}

In order to illustrate the debate about how informative are average flows, we study a simple two-dimensional linear stochastic system
\begin{align}
    \frac{\mathrm{d}y_t}{\mathrm{d}t} =& J_{yy} y_t  + J_{yb} b_t +  \omega_{y,t},
    \\\frac{\mathrm{d}b_t}{\mathrm{d}t} =& J_{by} y_t  + J_{bb} b_t +  \omega_{b,t},
    \label{eq-app:langevin-dynamics}
\end{align}
which can be studied as the Langevin dynamics described in \ref{app:linear-langeving-dynamics}. We select a variance of $\omega_{y,t},\omega_{b,t}$, $\varsigma^2=1$, and
\begin{align}
\*J = 
\begin{bmatrix}
   -1 & 1.2
  \\ -1  & -1
\end{bmatrix}.
\end{align}
The resulting covariance matrix is
\begin{align}
\^\Sigma = 
\begin{bmatrix}
   0.9545 &0.04545
  \\0.04545 &1.05455
\end{bmatrix}.
\end{align}
 Fig.~\ref{fig:2D-timeseries} displays an example of a trajectory of this system.

\subsection{Conditional marginal flows capture partial tendencies, not real behaviour}

Flows in the system are described as
\begin{align}
     f_y(y_t,b_t) =&  J_{yy} y_t + J_{yb} b_t,
     \\ f_b(y_t,b_t)  =& J_{by} y_t + J_{bb} b_t.
\end{align}

Following the results in \ref{app:linear-langeving-dynamics}, conditional average flows can be described as
\begin{align}
     \ang{f_y(y_t,b_t)}_{b_t} =&  J_{yy} m_y(b_t) + J_{yb} b_t
     \nonumber\\ =& \pr{J_{yy} \Sigma^*_{yb} /\Sigma^*_{bb} + J_{yb} }b_t,
     \label{eq-app:marginal-flows}
     \\ \ang{f_y(y_t,b_t)}_{y_t} =& \pr{J_{yy} + J_{yb} \Sigma^*_{by} /\Sigma^*_{yy} }y_t,
     \\ \ang{f_b(y_t,b_t)}_{b_t} =& \pr{J_{by} \Sigma^*_{yb} /\Sigma^*_{bb} + J_{bb} }b_t,
     \\ \ang{f_b(y_t,b_t)}_{y_t}  =& \pr{J_{by}  + J_{bb} \Sigma^*_{by} /\Sigma^*_{yy} }y_t.
\end{align}

\begin{figure}
\begin{center}
\begin{tabular}{ccc}
    \multicolumn{1}{l}{\textbf{A}}  & \multicolumn{1}{l}{\textbf{B}}  & \multicolumn{1}{l}{\textbf{C}}   \\
     \includegraphics[width=4.2cm]{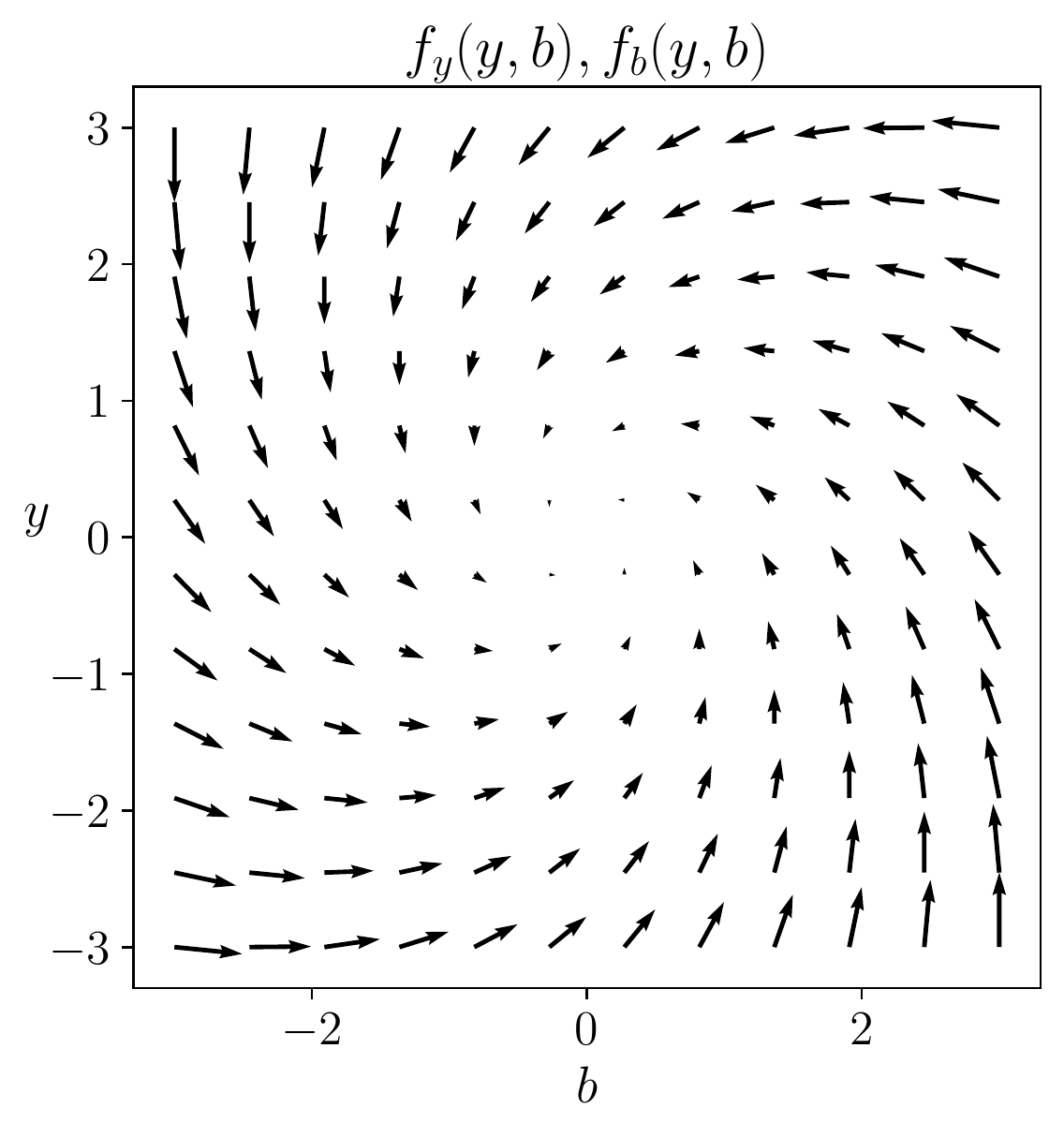} 
     &  \includegraphics[width=4.2cm]{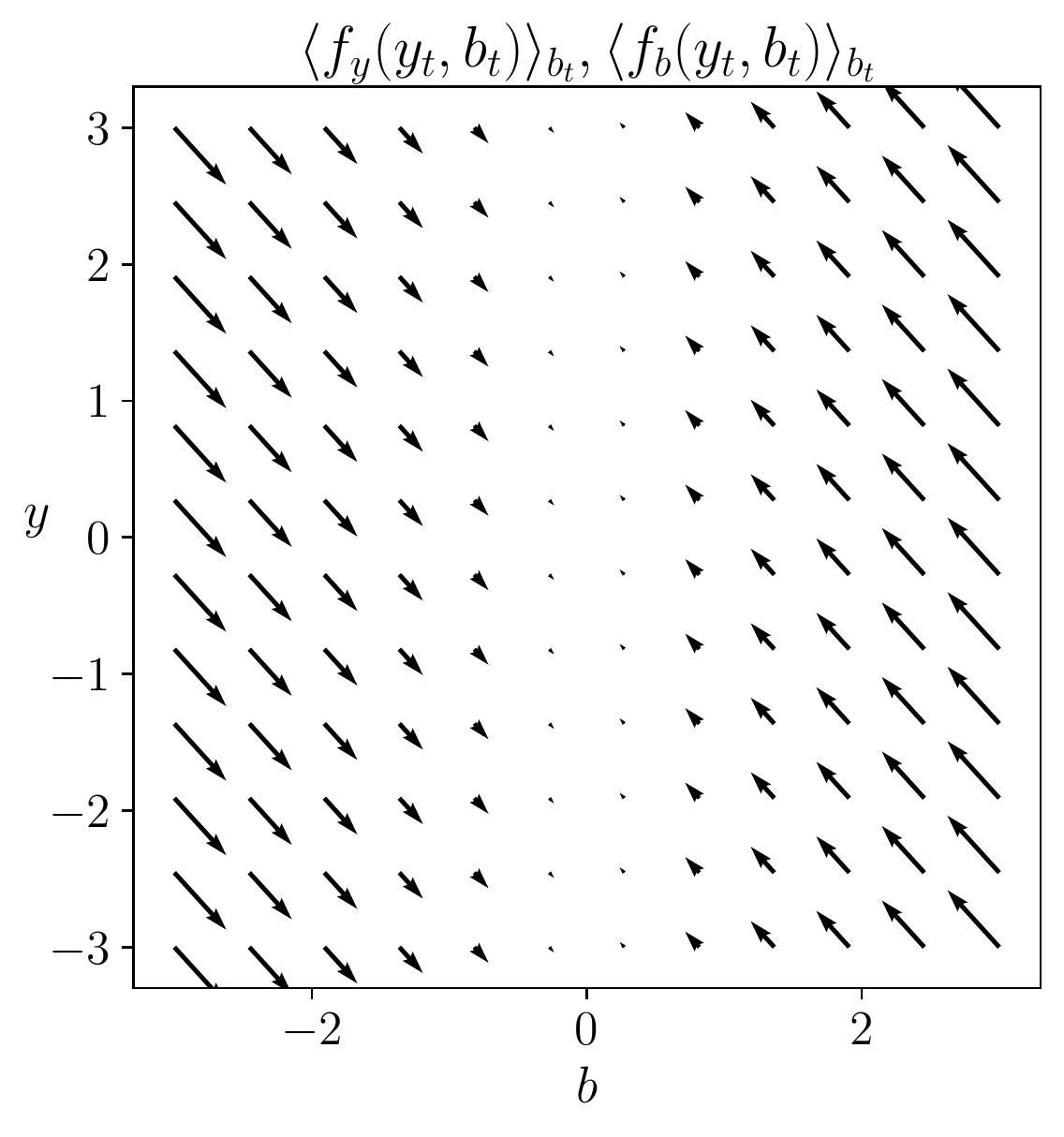}
     &  \includegraphics[width=4.2cm]{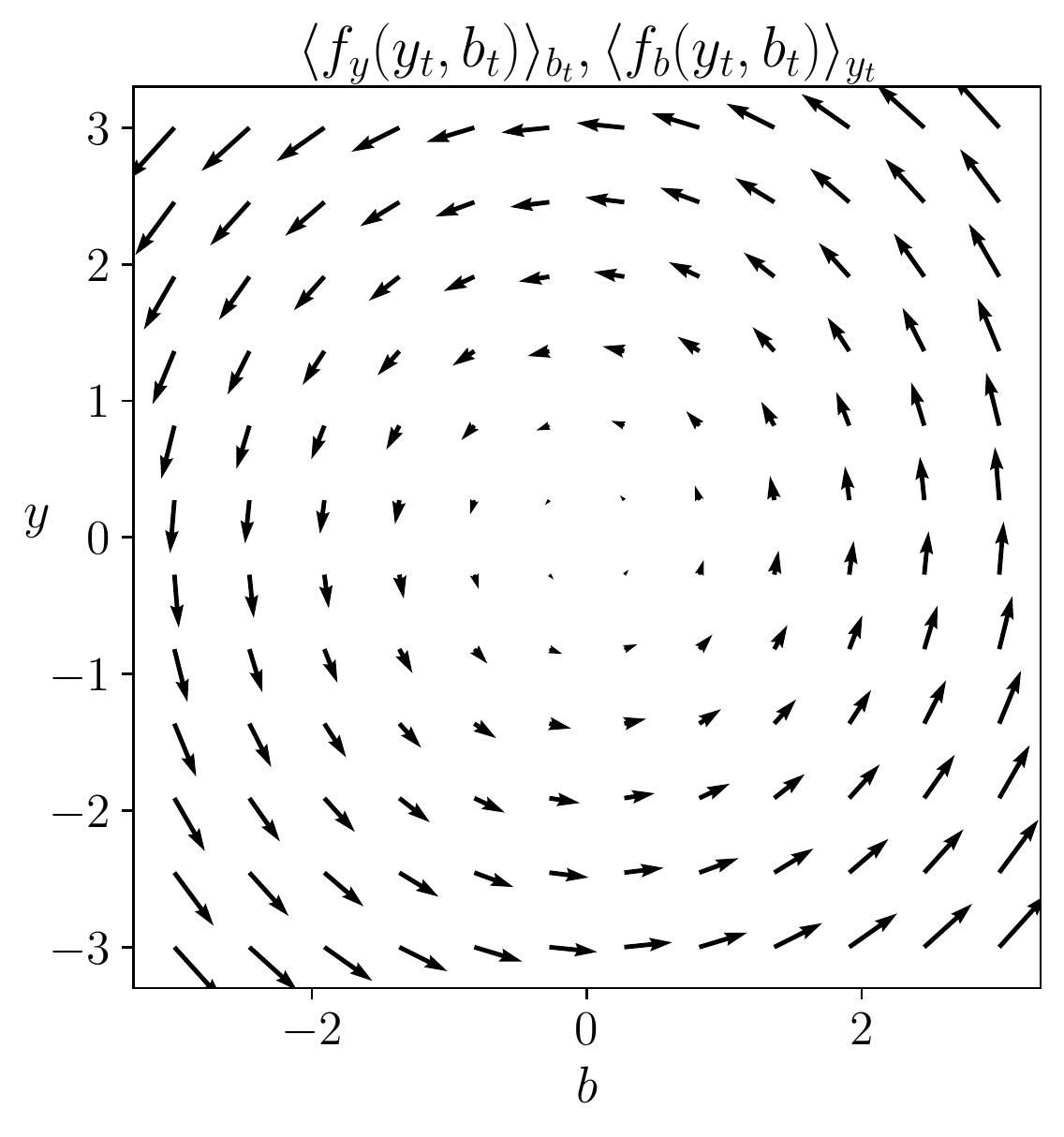}
\end{tabular}
\end{center}
\caption{\textbf{Flow structure vs conditional average flow structure of the system}. 
Comparison between the flow structure in the system displaying a spiral behaviour (\textbf{A}), the conditional average flows when variable $b$ is fixed, displaying divergent tendencies in $y$ and convergent tendencies in $b$ (\textbf{B}), and the cross conditional average flow when variable $b$ is fixed for $f_y$ and variable $y$ is fixed for $f_b$, displaying rotation without an attractor (\textbf{B}).
} 
 \label{fig:2D-flows}
 \end{figure}

In Fig.~\ref{fig:2D-flows}.A we describe the flow structure in the system, which describes a spiral behaviour due to the non-equilibrium tendencies in the system. This is an example of the solenoidal flows captured by the matrix $\*Q$.

In contrast, Fig.~\ref{fig:2D-flows}.B captures the marginal flow structure when variable $b$ is fixed. This separates tendencies in the system, resulting in a diverging tendency for variable $y$ (diverging to $-\infty$ for negative $b$ and to $+\infty$ for positive $b$.
Similarly, combining cross marginal flows (fix $b$ for the marginal flow of $y$ and vice versa) result in another combination of partial tendencies as shown in  Fig.~\ref{fig:2D-flows}.C. In this case, the rotation of solenoidal flows is captured, but no the attractor that structures this rotation into an spiral behaviour.

In sum, conditonal marginal flows do not capture the behaviour of the system. Even if the flows point in some direction (e.g. free energy minimization) we cannot conclude that this is equivalent to the system behaving `as if' following that particular direction.
An alternative illustration of this problem can be described by writing the conditional average flow of $y$ in terms of the most likely state $m_y(b)$
\begin{align}
     \ang{f_y(y_t,b_t)}_{b_t} =& \pr{J_{yy}  + J_{yb} \Sigma^*_{by} /\Sigma^*_{yy}} m_y(b_t).
     \label{eq-app:marginal-flows2}
\end{align}
Where the term $\pr{J_{yy}  + J_{yb} \Sigma^*_{by} /\Sigma^*_{yy}}$ is a positive constant (see Fig.~\ref{fig:2D-avrates}.B).
Following the logic of a relaxed interpretation of the FEP, this could be interpreted as a gradient ascent on $m_y^2(b_t)$. However, as we see in the Fig.~\ref{fig:2D-flows}.A the real behaviour of the system will display a global attractor at $y=b=0$ with solenoidal couplings, eventually minimizing  $m_y^2(b_t)$.

\subsection{Rates of conditional averages are different to conditional marginal flows}

The second issue we illustrate in this simple system is the difference between conditional marginal flows and the dynamics of the most likely states.

From \ref{app:linear-langeving-dynamics} we can derive
\begin{align}
     \frac{\mathrm{d}m_y(b_t)}{\mathrm{d}t} =& {\Sigma}^*_{yb} /{\Sigma}^*_{bb} \frac{\mathrm{d}b_t}{\mathrm{d}t}
	 = {\Sigma}^*_{yb} /{\Sigma}^*_{bb} \pr{ J_{by}y_t + J_{bb}b_t + \varsigma \omega_{b,t}}.
	\label{eq-app:most-likely-state-evolution}
\end{align}
For a fixed $b$, this variable is distributed as a Normal distribution $\mathcal{N}(u,s)$, with mean and standard deviation
\begin{align}
    u=&{\Sigma}^*_{yb} /{\Sigma}^*_{bb}  J_{by} m_y(b),
    \\s =& {\Sigma}^*_{yb} /{\Sigma}^*_{bb} \pr{J_{by} (\Sigma^*_{yy} - \Sigma^*_{yb}\Sigma^*_{by} /\Sigma^*_{bb} ) + \sqrt{2} \varsigma s_\omega},
\end{align}
where $s_\omega$ is the standard deviation of the noise introduced in the Langevin dynamics (described as $\sqrt{\frac{1}{\mathrm{d}t}}$).

\begin{figure}
\begin{center}
\begin{tabular}{cc}
    \multicolumn{1}{l}{\textbf{A}}  & \multicolumn{1}{l}{\textbf{B}}    \\
     \includegraphics[width=6.8cm]{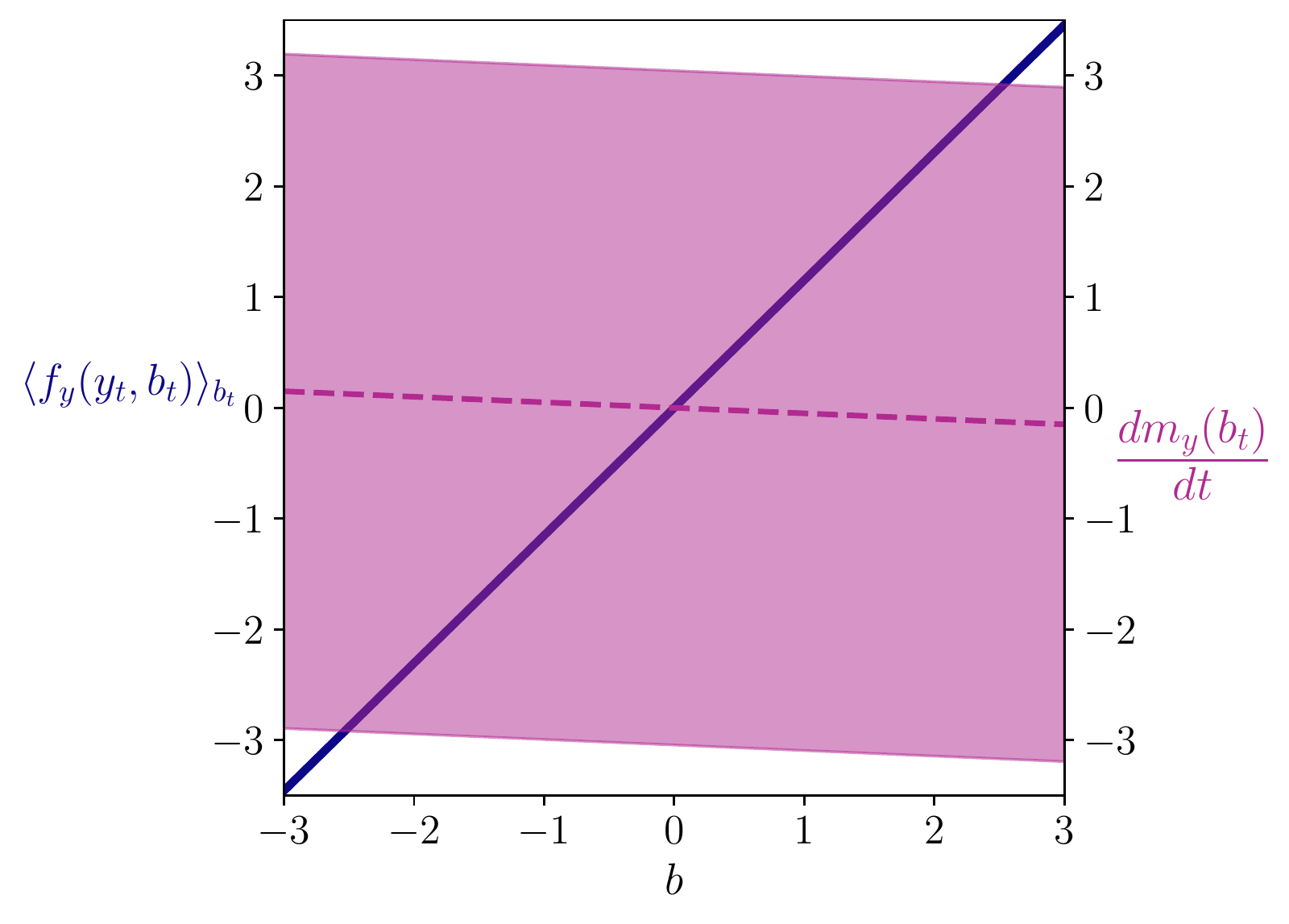} 
     &  \includegraphics[width=6.8cm]{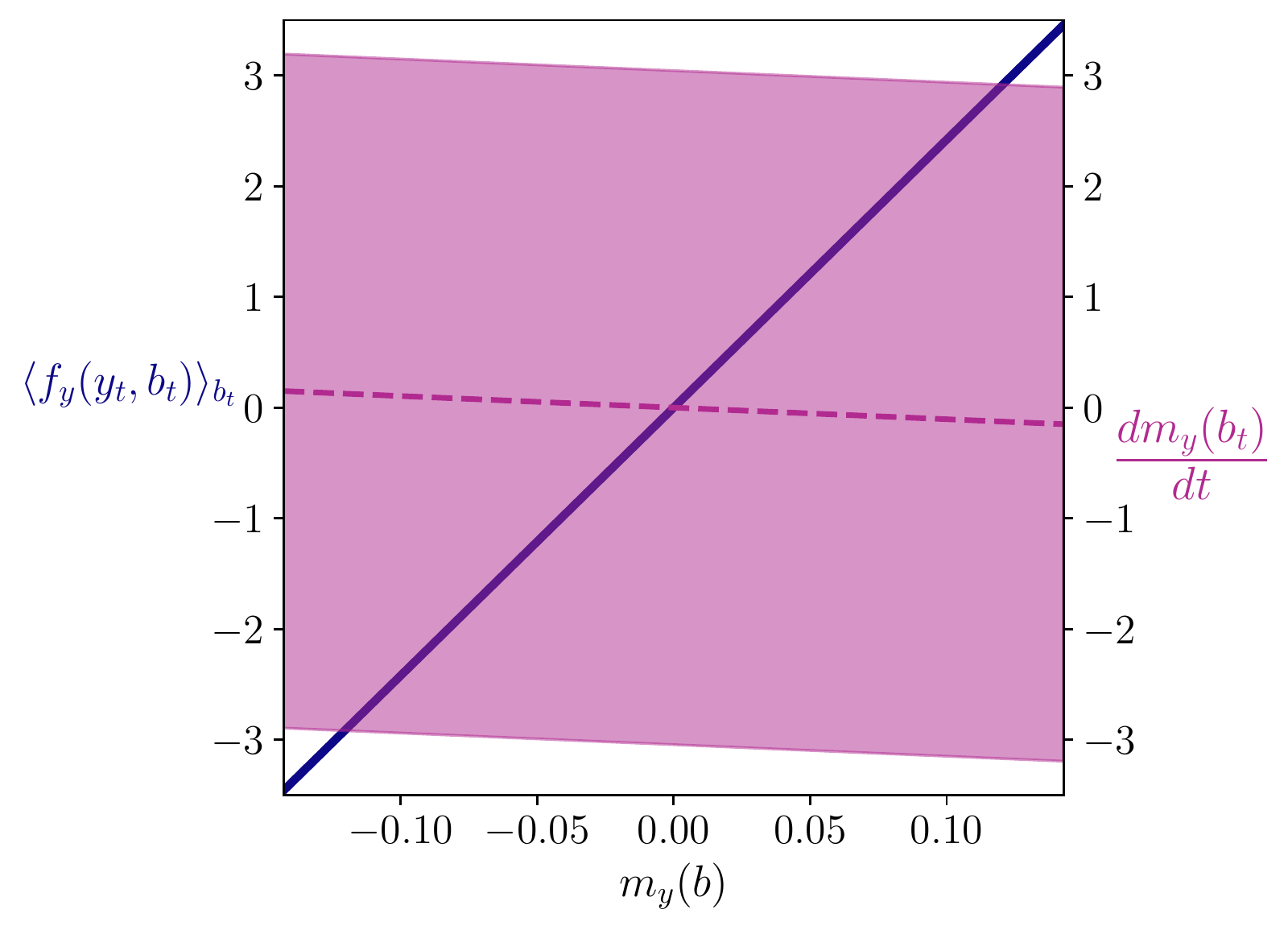}
\end{tabular}
\end{center}
\caption{\textbf{Rate of conditional averages versus conditional average flows}. The conditional average flow of $y$ fixing $b$, $\ang{f_y(y_t,b_t)}_{y_t}$, is represented by the solid dark line. The rate of change of the conditional average $\frac{m_y(b_t)}{\mathrm{d}t}$ is represented by the dashed line (representing its average value for a given $b$, and the light area (representing error bars for 3 standard deviations for a fixed $b$.). These quantities are represented with respect to a $b$ axis (\textbf{A}) and a $m_y(b)$ axis (\textbf{B}), as one is a linear transformation of the other.
} 
 \label{fig:2D-avrates}
 \end{figure}

For a range of values of $b$, Fig.~\ref{fig:2D-avrates} captures the conditional average flows $\ang{f_y(y_t,b_t)}_{b_t}$ (solid dark line), and the distribution of the derivatives of the conditional average state $\frac{\mathrm{d}m_y(b_t)}{\mathrm{d}t} $ (the mean is represented by the dashed line and error bars of three standard deviations by the light area).
As we can observe, the dependency with $m_y(b)$ is positive with respect to $\ang{f_y(y_t,b_t)}_{b_t}$ (indicating a gradient ascent on $m_y^2(b)$) while the true dynamics is captured by  the negative dependency between $\frac{\mathrm{d}m_y(b_t)}{\mathrm{d}t} $ and $m_y(b)$ (indicating a gradient descent on $m_y^2(b)$).
Note that for different parameters the sign of the slope of  $\ang{f_y(y_t,b_t)}_{b_t}$ can change (being negative or positive) but the slope of $\frac{\mathrm{d}m_y(b_t)}{\mathrm{d}t} $ is always negative, given the presence of a global attractor. This shows how, even in very simple examples, these quantities can have radically different behaviours and that the conditional average flow does not necessarily capture the true behaviour of a system.

\end{document}